\documentclass[final,twocolumn,openright]{pasj01}
\usepackage{textcomp}
\usepackage{threeparttable}
\usepackage[]{graphicx}
\Received{}
\Accepted{}
 
\begin{document} 
\raggedbottom
\title{Massive core/star formation triggered by cloud-cloud collision: Effect of magnetic field}

\author{Nirmit \textsc{Sakre}\altaffilmark{1,*}%
}
\altaffiltext{1}{Department of Physics, Faculty of Science, Hokkaido University, Kita 10 Nishi 8, Kita-ku, Sapporo, Hokkaido 060-0810, Japan}
\email{nirmit@astro1.sci.hokudai.ac.jp}

\author{Asao \textsc{Habe}\altaffilmark{1}}

\author{Alex R. \textsc{Pettitt}\altaffilmark{1},}
\author{Takashi \textsc{Okamoto}\altaffilmark{1}}

\KeyWords{ISM: clouds --- ISM: magnetic fields --- stars: formation --- stars: massive}

\maketitle

\begin{abstract}
We study the effect of magnetic field on massive dense core formation in colliding unequal molecular clouds by performing magnetohydrodynamic simulations with sub-parsec resolution (0.015 pc) that can resolve the molecular cores. Initial clouds with the typical gas density of the molecular clouds are immersed in various uniform magnetic fields. The turbulent magnetic fields in the clouds consistent with the observation by \citet{2010ApJ...725..466C} are generated by the internal turbulent gas motion before the collision, if the uniform magnetic field strength is 4.0 $\mu$G. The collision speed of 10 km s$^{-1}$ is adopted, which is much larger than the sound speeds and the Alfv\'{e}n speeds of the clouds. We identify gas clumps with gas densities greater than 5 $\times$ 10$^{-20}$ g cm$^{-3}$ as the dense cores and trace them throughout the simulations to investigate their mass evolution and gravitational boundness. We show that a greater number of massive, gravitationally bound cores are formed in the strong magnetic field (4.0 $\mu$G) models than the weak magnetic field (0.1 $\mu$G) models. This is partly because the strong magnetic field suppresses the spatial shifts of the shocked layer that should be caused by the nonlinear thin shell instability. The spatial shifts promote the formation of low-mass dense cores in the weak magnetic field models. The strong magnetic fields also support low-mass dense cores against gravitational collapse. We show that the numbers of massive, gravitationally bound cores formed in the strong magnetic field models are much larger than in the isolated, non-colliding cloud models, which are simulated for comparison. We discuss the implications of our numerical results on massive star formation.

\end{abstract}
\section{Introduction}
Massive stars have fundamental influence over the interstellar medium and galactic evolution. They ionize the surrounding gas and deposit energies by strong stellar winds and supernovae. Massive stars supply new material to the surrounding gas, which is available for next-generation stars. Massive star formation still remains poorly understood despite decades of work \citep{2007ARA&A..45..481Z,2014prpl.conf..149T}. The accretion rate, $\dot M$, onto a protostar is given by,
\begin{equation}
\dot M \sim \frac{c_T^3}{G}\sim 10^{-6} \left(\frac{T}{10 K}\right)^{3/2} M_{\odot} \textrm{ yr}^{-1},
\end{equation}
where $c_T=(k_{\rm B} T/m)^{1/2}$ is the isothermal sound speed of gas of mean molecular mass, $m$, at gas temperature, $T$, $k_{\rm B}$ is Boltzmann constant, and G is the gravitational constant \citep{1977ApJ...214..488S, 1980ApJ...241..637S}. \citet{1980ApJ...241..637S} proposed that a low-mass star is formed in a molecular core with $T$ = 10 K.
\citet{2002Natur.416...59M} proposed that a higher rate of gas accretion on to a protostar is needed to form a massive star in a molecular core, since ram pressure associated with the accretion can exceed the radiation pressure from the protostar. 
Such a high accretion rate can be realized in a massive dense core with large turbulence or high Alfv\'en wave speed. The study of the formation and evolution of such massive dense cores remains an integral part of understanding the massive star formation process.
 
Cloud-cloud collisions (CCCs) are strong candidates for massive star formation.
If two molecular clouds collide at supersonic speed, a shock wave is produced at the interface of the colliding clouds. Since the density in the shocked region increases further due to radiative cooling, dense clumps can be formed due to the enhanced self-gravity \citep{1984ApJ...279..335G}.
If the sizes of the colliding clouds are different, a converging flow appears in the shocked layer, and it can increase the mass of the dense clumps \citep{1992PASJ...44..203H}.
If the dense cores formed in the massive clump are massive enough, we can expect massive star formation in them.
 
Observational evidence of massive star formation by CCCs are reported in various star formation regions (e.g., \citet{1994ApJ...429L..77H,2009ApJ...696L.115F,2010ApJ...709..975O,2016ApJ...819...66D,2017ApJ...835..142T,2018PASJ...70S..46F,2018ApJ...859..166F}). These are super star clusters, the H$_{\mathrm{II}}$ regions, the Spitzer bubbles in the Milky Way disk, and star formation regions in the Large Magellanic Cloud. Observational evidence consists of two distinct velocity components of molecular gas with a large velocity difference (more than 10 km s$^{-1}$) and bridge features in the position-velocity diagrams of observed molecular gas in these regions. These features were found in recent numerical simulations of CCCs \citep{2014ApJ...792...63T,2015MNRAS.454.1634H,2015MNRAS.450...10H,2018PASJ...70S..58T}. The typical collision speeds observed in these regions are in the range of 10 to 20 km s$^{-1}$, which is much higher than the gas speeds due to internal motion in molecular clouds. 

Recently, hydrodynamic simulations of CCCs have been carried out to study the formation of dense cores, assuming turbulence in the clouds  \citep{2014ApJ...792...63T,2018PASJ...70S..58T,2018PASJ...70S..54S}. They discuss that dense cores formed in the shocked layer can increase their mass by accretion of dense gas if turbulence and magnetic fields support the dense cores. Simulations with greater spatial resolution are needed to study the effect of turbulence and magnetic fields on dense core formation.

Magnetohydrodynamic (MHD) simulations of colliding molecular clouds have been performed by several authors \citep{2013ApJ...774L..31I,2014ApJ...785...69C,2015ApJ...810..126C,2017ApJ...841...88W,2017ApJ...835..137W,2018PASJ...70S..53I}. 
\citet{2014ApJ...785...69C} and \citet{2015ApJ...810..126C} have shown that the magnetic fields do not affect the typical mass of pre-stellar cores formed in the colliding clumps in a large molecular cloud.
They found the formation of low-mass stars in their simulations. They selected collision speed as large as the typical turbulent velocity observed in molecular clouds. 
Since the observed CCC speeds are much larger than the typical turbulent velocity, there is an obvious need to study the collision of magnetized clouds with collision speeds more typical of those observed. 
\citet{2017ApJ...841...88W} and \citet{2017ApJ...835..137W} have shown that the magnetic fields do not affect star formation in the CCC in their numerical simulations with more typical collision speed of CCCs, and they investigated observational signatures and physical properties of dense regions in the colliding magnetized clouds.
In their simulations, they used a spatial resolution of $\sim$ 0.1 pc that may not be high enough to resolve the dense cores of which typical scale is $\sim$ 0.1 pc \citep{2007ARA&A..45..339B}, and a higher spatial resolution is required to properly reproduce their physical properties. \citet{2011ApJ...731...62F} proposed a minimum resolution criterion of 30 cells per Jeans length in hydrodynamic simulations of self-gravitating gas in order to resolve turbulent motion on the Jeans scale. A similar number of cells should be used to resolve the dense cores.
We use a minimum cell size of 0.015 pc after some numerical tests using a different choice of minimum cell size (see subsection \ref{numerical_methods}) and define dense cores using a selection criteria in line with the typical density in observed dense cores, described in detail in sub-subsection \ref{dense-core}. Simulations of the collision of magnetized clouds with typical speed and sufficient spatial resolution were performed by \citet{2013ApJ...774L..31I} and \citet{2018PASJ...70S..53I}.  
Their results are in favor of massive star formation due to the role of magnetic fields. \citet{2018PASJ...70S..53I} simulated a collision of dense regions in a giant molecular cloud (GMC), assuming a uniform initial magnetic field 20 $\mu$G perpendicular to the collision velocity.
Though the studies mentioned above have reached sufficient resolutions and have used realistic collision speeds, there is a dearth of studies that have investigated this while considering alternative magnetic field configurations (both in strength and orientation with respect to the CCC axis) and the typical GMC-scale CCCs.

We present a study of the effects of magnetic fields on the formation of massive dense cores in magnetized, turbulent, and colliding clouds using high spatial resolution simulations. We perform simulations assuming magnetic fields of varying strengths and directions to understand the role of magnetic fields on massive dense core formation. In section \ref{methods} we describe the numerical method and models, in section \ref{results} we present our numerical results, in section \ref{discussion} we discuss our results, and in section \ref{summary} we summarize our study.
\section{Numerical method and models}\label{methods}
\subsection{Numerical method}\label{numerical_methods}
We use simulation code $Enzo$, a three-dimensional MHD adaptive mesh refinement (AMR) code \citep{2014ApJS..211...19B}. We assume ideal MHD in our simulations. The code solves the MHD equations using the MUSCL 2nd-order Runge-Kutta temporal update of the conserved variables with the Harten-Lax-van Leer (HLL) method and a piecewise linear reconstruction method (PLM). The hyperbolic divergence cleaning method of \citet{2002JCoPh.175..645D} is adopted to ensure the solenoidal constraint on the magnetic field.

We describe numerical methods of the cooling, the pressure floor, and the Alfv\'{e}n speed limiter used in our simulations. Radiative cooling of gas is calculated down to 10 K by using the cooling table made by the CLOUDY cooling code \citep{1998PASP..110..761F} with the solar metallicity and a density $n_{\rm H}$ = 100 cm$^{-3}$. For example, the cooling time of gas with a density $n_{\rm H}$ = 100 cm$^{-3}$ from 100 K to 10 K is estimated to be less than 0.1 Myr by using the cooling table. Due to this rapid cooling of molecular gas, the dense gas reaches the typical cloud temperature of GMC, 10 K. Photoelectric heating with the rate of 1.2 $\times$ 10$^{-25}$ (n$_{\rm H}$/1 cm$^{-3}$) erg s$^{-1}$ cm$^{-3}$ is applied to gas \citep{2008ApJ...673..810T}. Self-gravity in the gas is included in our simulations. The pressure floor is applied for the cell in the finest grid level in which absolute value of self-gravitational energy of gas is greater than its internal energy \citep{2001ApJ...548..509M}. The pressure floor kicks in at a gas density of $\sim$ 10$^{-15}$ g cm$^{-3}$ for a cell at gas temperature of 10 K in the finest grid level. Tests with a higher (lower) value of the pressure floor parameter by a factor of 10 (1/3) resulted in a very similar core population and probability density functions, and so our simulation results are unlikely to be influenced by any numerical effects of our default pressure floor parameter. The Alfv\'{e}n speed, $v_{\rm A}$, is given by
\begin{equation}
v_{\rm A}=\frac{B}{\sqrt{4\pi\rho}},
\end{equation}
where $B$ is magnetic field strength, $\rho$ is density, and the CGS system of units are used in all equations related to magnetic field in this paper.
The maximum Alfv\'{e}n speed is set to be 20 km s$^{-1}$ to avoid very short time-steps created due to high Alfv\'{e}n speeds in low gas density regions, which is effectively limited by increasing the gas density in such regions. We find that the Alfv\'{e}n speed limiter works in very small regions with much lower gas density than the initial density of our model clouds.
\begin{table}
\caption{Initial cloud model parameters.}
\label{tab:cloud}
\centering
    \begin{threeparttable}
    \centering
    \resizebox{\columnwidth}{!}{%
    \begin{tabular}{  c  c  c  c  c }
    \hline
    Parameter\tnote{*} & Isolated cloud & Small cloud & Large cloud  & Units \\ \hline
    $R$ & 7.3 & 3.5  & 7 & pc  \\ 
    $M$ & 8746 & 972 & 7774   & $M_{\odot}$ \\ 
    $\rho_{0}$ &3.67 $\times$10$^{-22}$   & 3.67 $\times$10$^{-22}$ &  3.67 $\times$10$^{-22}$ & g cm$^{-3}$ \\ 
    
    $t_{\rm ff}$ & 3.5 & 3.5 & 3.5 & Myr \\ 
    $\sigma_{v}$ & 2.1 & 1.0 & 2.0  & km s$^{-1}$ \\ 
    $v_{\rm coll}$ & --- & 10  & 0 & km s$^{-1}$   \\ \hline 
    \end{tabular}%
}
    \begin{tablenotes}
    \item[*]$R$---cloud radius, $M$---cloud mass, $\rho_{0}$---cloud initial\\ density, $t_{\rm ff}$---free-fall time of the cloud, $\sigma_{v}$---the \\velocity dispersion of the cloud, $v_{\rm coll}$---the collision\\ speed of the cloud.
    \end{tablenotes}
    \end{threeparttable}
\end{table}
The minimum density in our simulations is selected as the initial density of the ambient medium of 1.69$\times$10$^{-23}$ g cm$^{-3}$, as given in sub-subsection \ref{2_2_1section}. Tests with a higher value of the maximum Alfv\'{e}n speed by a factor of 5 and with a lower value of the minimum density by a factor of 1/10 resulted in a very similar core population and probability density functions to our simulation results with our default values of the maximum Alfv\'{e}n speed and the minimum density. We use yt, a multi-code analysis toolkit for astrophysical simulation data, \citep{2011ApJS..192....9T} to analyze our numerical results. The yt is very powerful to analyze numerical results given by $Enzo$ code.

\subsection{Cloud models}\label{2_2section}
\subsubsection{Initial cloud structure and collision setup}\label{2_2_1section}
We adopt initial conditions for clouds based on properties of observed GMCs \citep{2009ApJ...699.1092H,2011ApJ...729..133M}. Two uniform molecular clouds, a small cloud and a large cloud, are initialized with density 3.67 $\times $10$^{-22}$ g cm$^{-3}$ of which free-fall time is 3.5 Myr and with masses 972 $M_{\odot}$ and 7774 $M_{\odot}$, respectively. We stop our simulations at $t$ = 3.0 Myr, which is earlier than the free-fall time.
These cloud masses are rather small in comparison to observed GMCs in the Milky Way of which the mass range is 10$^3$-10$^6$ $M_{\odot}$ \citep{2011ApJ...729..133M}. We select small clouds to achieve high spatial resolution needed to study the effect of the magnetic field on the formation of massive dense cores in the colliding clouds with a rather small simulation box that we describe in the next paragraph. 
We adopt initial temperatures of the clouds as 68 K and 273 K for the small cloud and the large cloud, respectively.
While such temperatures are sufficient to provide initial pressure support, the dense gas in the clouds rapidly cools down to 10 K due to the radiative cooling during evolution.
We adopt turbulence in clouds (see sub-subsection \ref{magnetic-field}).
Parameters for each cloud are summarized in table \ref{tab:cloud}. 
A typical collision speed of 10 km s$^{-1}$ is used.
The ambient medium has a density of 1.69 $\times$ 10$^{-23}$ g cm$^{-3}$ and a temperature of 800 K.
This high density of the ambient medium is used to avoid high Alfv\'{e}n speeds in the ambient medium.

\begin{table}
\caption{Simulation models.}
\label{tab:model}
\centering
    \begin{threeparttable}
    \centering
    \resizebox{0.8\columnwidth}{!}{%
    \begin{tabular}{  c  c  c  c }
    \hline 
    Model no. & Model name & ${B_0}$ ($\mu$G)\tnote{*} & $\theta$\tnote{$\dagger$} \\ \hline 
    1 &  Xweak & 0.1 & $0^\circ $  \\ 
    2 & Yweak & 0.1 & $90^\circ $  \\ 
    3 & XYweak & 0.1 & $45^\circ $ \\ 
    4 & Xstrong & 4.0 & $0^\circ $  \\ 
    5 &  Ystrong & 4.0 & $90^\circ $  \\ 
    6 &  XYstrong & 4.0 & $45^\circ $ \\ 
    7 &  ISweak & 0.1 & $90^\circ $ \\ 
    8 &  ISstrong & 4.0 & $90^\circ $ \\ \hline 
    \end{tabular}%
}
    \begin{tablenotes}
    \item[*]The initial magnetic field strength, 
    \item[$\dagger$]The angle between the initial magnetic field, $\boldsymbol B_0$, and the collision axis ($x$-axis).
    \end{tablenotes}
    \end{threeparttable}
\end{table}

Six different colliding clouds models are considered, each with differing initial magnetic field strength and direction (see sub-subsection \ref{magnetic-field} and table \ref{tab:model}). 
Our simulation domain encompasses (32 pc)$^3$ with root grids 128$^3$, and we use four refinement levels based on the condition of minimum baryon mass of 0.05 $M_{\odot}$ for refinement. This gives the minimum cell size of 0.015 pc at the maximum refinement level. We have tested our simulations with additional higher refinement levels; level five and level six. The minimum cell size is 0.0075 pc for the refinement level five and 0.0037 pc for the refinement level six. The minimum cell size of 0.0037 pc satisfies the minimum resolution criterion in hydrodynamic simulations of self-gravitating gas proposed by \citet{2011ApJ...731...62F} for the typical core size of 0.1 pc. We find that core mass functions and core properties in simulations with these higher refinement levels are very similar to the simulation results with our default refinement levels.
  
Additionally, two different isolated, non-colliding cloud models are used for comparison with the results of colliding clouds models (e.g., the population of dense cores). This isolated cloud has a sum of the masses of both the small and the large clouds, and it also includes turbulent motion (see table \ref{tab:cloud}). Two different initial magnetic field strengths are selected similarly to the colliding clouds (see sub-subsection \ref{magnetic-field}).
We summarize the simulation results of the isolated cloud models in the Appendix.   

\subsubsection{Magnetic field and turbulence in clouds}\label{magnetic-field}
The clouds are immersed in an initial uniform magnetic field, $\boldsymbol B_0$, and turbulent motions develop inside them from $t$ = 0 to 0.5 Myr. A turbulent magnetic field is generated inside clouds before the small cloud begins to move towards the large cloud at $t$ = 0.5 Myr along the collision axis (positive $x$-axis of the simulation box). 
We select two strengths of $\boldsymbol B_0$, $B_0$ = 0.1 $\mu$G (weak) and $B_0$ = 4.0 $\mu$G (strong) and three directions of $\boldsymbol B_0$, which are parallel to the collision axis, perpendicular (along positive $y$-axis) to the collision axis, and oblique to the collision axis. The angle, $\theta$, between $\boldsymbol B_0$ direction and collision axis for oblique $B_0$ model is 45$^\circ$. Additional isolated cloud models have $\boldsymbol B_0$ with magnetic field strengths $B_0$ = 0.1 $\mu$G (weak) and 4.0 $\mu$G (strong) similar to those selected for colliding cloud models. The direction of $\boldsymbol B_0$ is taken along the positive $y$-axis of the simulation box. We name the simulation models as shown in table \ref{tab:model}.

Turbulent velocities are generated to be consistent with the Larson relation \citep{1981MNRAS.194..809L, 2009ApJ...699.1092H} at $t$ = 0 Myr, by imposing a velocity field with power spectrum ${v_k}^{2}$ $\propto$ $k^{-4}$. We define $\boldsymbol{k}$ = $(2\pi)(n_x\boldsymbol{e_x}$ + $n_y\boldsymbol{e_y}$ + $n_z\boldsymbol{e_z}$)/($a R$), where $\boldsymbol{e_x}$, $\boldsymbol{e_y}$, and $\boldsymbol{e_z}$ are unit vectors in $x$-direction, $y$-direction, and $z$-direction, respectively, $R$ is the radius of each cloud, and $a$ $\sim$ 4. We use integers for $n_{x}$, $n_{y}$, and $n_{z}$ as $n_{\rm min}^2$ $\leq$ $n_x^2$ + $n_y^2$ + $n_z^2$ $\leq$ $n_{\rm max}^2$, where $n_{\rm min}$ and $n_{\rm max}$ are 6 and 12 for the small cloud and 8 and 15 for both the large cloud and isolated cloud, respectively. If the initial kinetic energy of turbulence is in virial equilibrium with the self-gravitational energy, the velocity dispersion, $\sigma_v$, of clouds due to the turbulent motion is $\sigma_v$ = 0.86 km s$^{-1}$ for the small cloud and $\sigma_v$ = 1.72 km s$^{-1}$ for the large cloud. We slightly increase these values to $\sigma_v$ = 1.0 km s$^{-1}$ for the small cloud and $\sigma_v$ = 2.0 km s$^{-1}$ for the large cloud to be consistent with the observational results by \citet{2009ApJ...699.1092H}. 

\begin{figure*}[!htb]
    \begin{center}
    \includegraphics[width=0.8\textwidth]{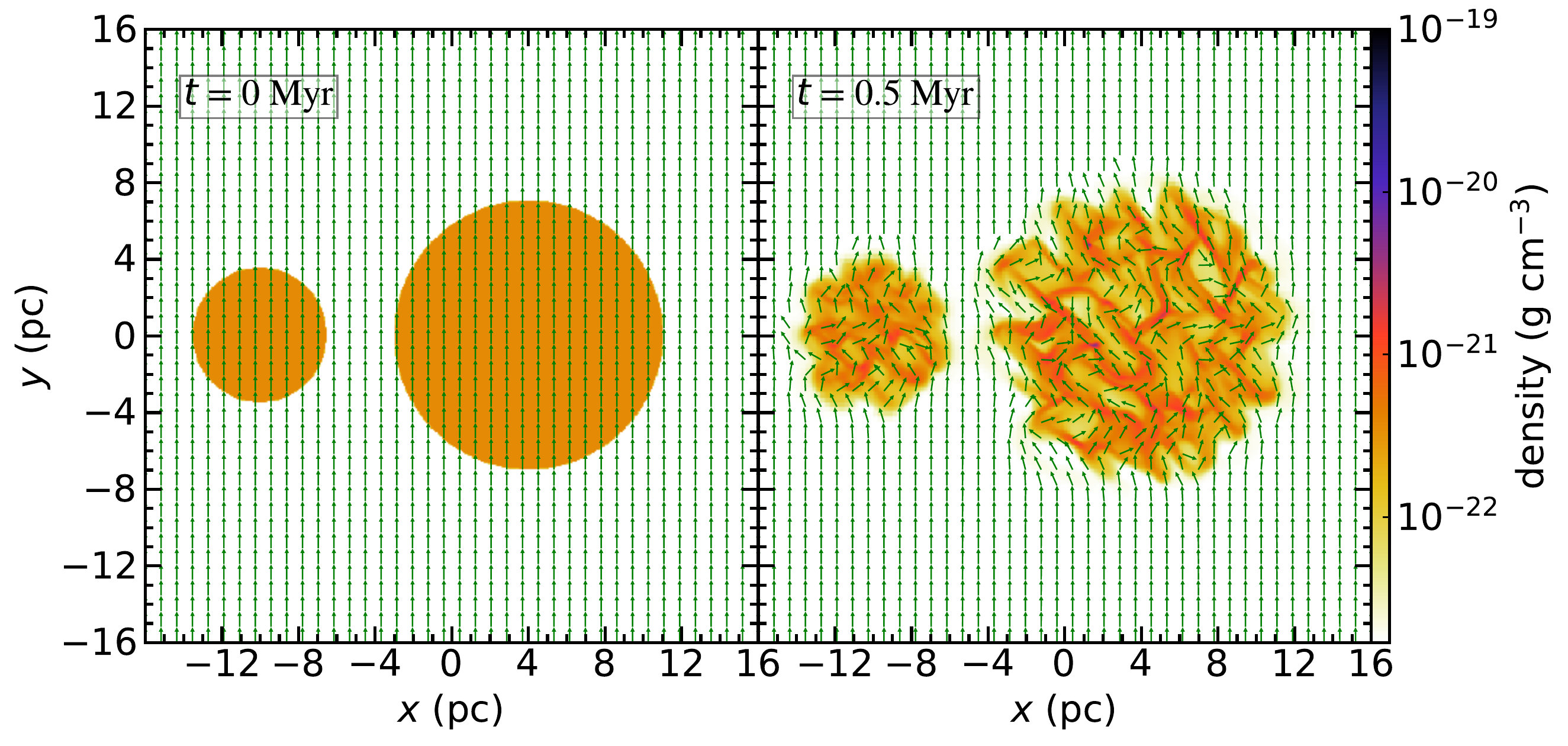}
    \end{center}
    \caption{Slice plots of the gas density in $z$ = 0 pc at $t$ = 0 Myr (left) and 0.5 Myr (right) for the Ystrong model. The arrows show normalized vectors, [$B_x$/(${B_x}^2$ + ${B_y}^2$)$^{1/2}$, $B_y$/(${B_x}^2$ + ${B_y}^2$)$^{1/2}$]. Color bar of the gas density is shown on the right edge.}
    \label{fig:1} 
\end{figure*}
\begin{figure*}[!htb]
    \begin{center}
    \includegraphics[width=0.8\textwidth]{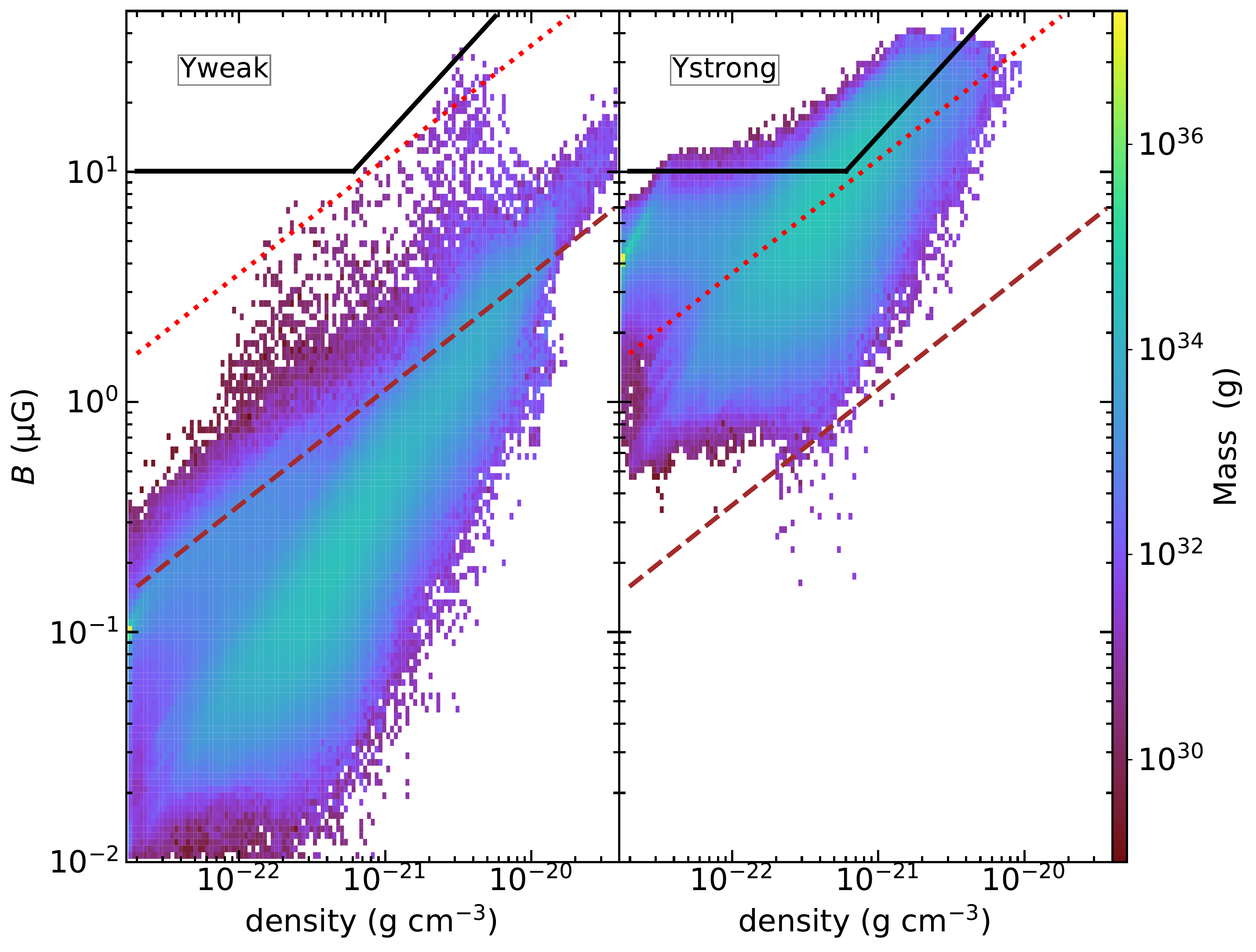}
    \end{center}
    \caption{Phase plot of the magnetic field strength and the gas density in the Yweak (left) and Ystrong (right) models at $t$ = 0.5 Myr. Solid lines show the observed relation by \citet{2010ApJ...725..466C}. Dotted lines show Alfv\'{e}n speed, $v_{\rm A}$ = 1 km s$^{-1}$ and dashed lines show Alfv\'{e}n speed, $v_{\rm A}$ = 0.1 km s$^{-1}$. The colors show total mass of the simulation cells with density, $\rho$, and magnetic field strength, $B$, in the range from $\rho$ to $\rho$(1 $+$ $\Delta_1$) in $x$-axis and $B$ to $B$(1 $+$ $\Delta_2$) in $y$-axis, where $\Delta_1$ = 0.063 and $\Delta_2$ = 0.069, and the color bar is shown on the right edge.}
    \label{fig:2} 
\end{figure*}
We show the effect of turbulence on the initial magnetic field before the collision starts. Figure \ref{fig:1} shows turbulent density structures and turbulent magnetic fields of the small and large clouds at $t$ = 0 and 0.5 Myr in the Ystrong model. 
In figure \ref{fig:1}, $\boldsymbol B_0$ is perpendicular to the collision axis.
The arrows in figure \ref{fig:1} show normalized vectors, [$B_x$/(${B_x}^2$ + ${B_y}^2$)$^{1/2}$, $B_y$/(${B_x}^2$ + ${B_y}^2$)$^{1/2}$]. The variation of the magnetic field direction in the clouds seen at $t$ = 0.5 Myr is due to the effect of turbulence.
We show the magnetic field and density relation in the simulation box of the Yweak (weak $B_0$) and the Ystrong (strong $B_0$) models at $t$ = 0.5 Myr in figure \ref{fig:2}.
In the Yweak model (left-hand panel of figure \ref{fig:2}), the turbulent magnetic fields in clouds are weaker than the observed relation between the magnetic field and gas density given by \citet{2010ApJ...725..466C}, indicated by a solid line in figure \ref{fig:2}. 
In the Ystrong model (right-hand panel of figure \ref{fig:2}), the turbulent magnetic fields in the clouds are consistent with the observed relation. Figure \ref{fig:2} shows that the magnetic fields in both clouds are dominated by the turbulent magnetic fields that are much stronger than $B_0$. We also show the constant Alfv\'{e}n velocities of 1 km s$^{-1}$ (dotted lines) and 0.1 km s$^{-1}$ (dashed line) in figure \ref{fig:2} for comparison with the turbulent velocities in clouds and the collision speed of the clouds, 10 km s$^{-1}$. The magnetic field and density relation before the collision are independent of the direction of $\boldsymbol B_0$. Hence, the pre-collision results of only the Ystrong and Yweak models are shown in this section.

\begin{figure*}
    \centering
    \includegraphics[width=0.45\textwidth]{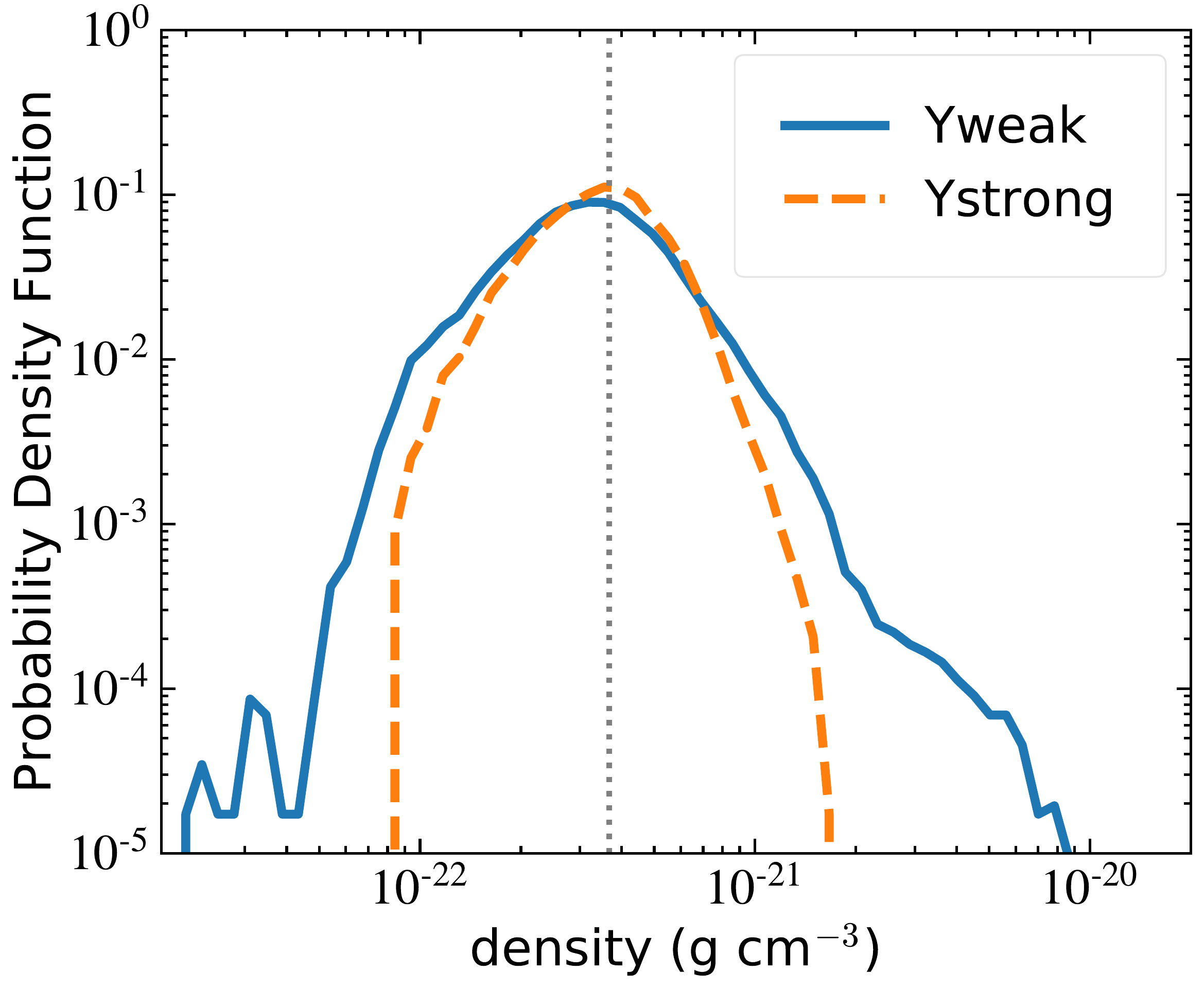}
    \includegraphics[width=0.45\textwidth]{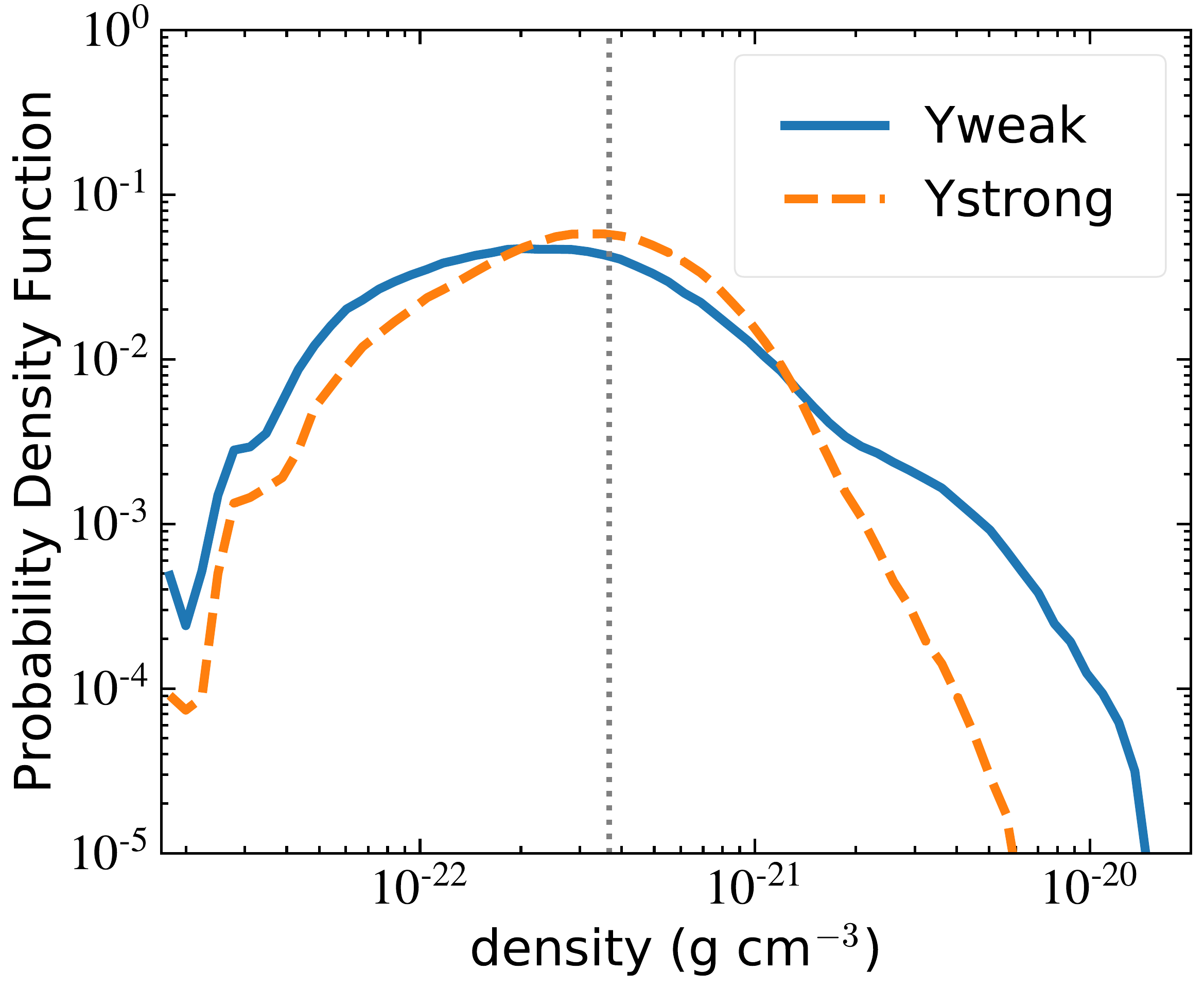}  
    \caption{Probability density functions (PDFs) of the small cloud (left) and the large cloud (right) at $t$ = 0.5 Myr for the Yweak (solid line) and Ystrong (dashed line) models. The vertical dotted line indicates the initial density of clouds. In order to neglect the effect of gas motion near each cloud surface, we select spheres of radii equal to 85 \% of the initial cloud radii centered at the initial cloud centers for PDFs.}
    \label{fig:3} 
\end{figure*}

The probability density functions (PDFs) of the small cloud and the large cloud at $t$ = 0.5 Myr before collision are shown in figure \ref{fig:3}. The initial density of both clouds is indicated by the
vertical dotted lines in figure \ref{fig:3}. The PDFs of both the clouds are log-normal shaped due to the effect of turbulence generated, and the extended tail is due to the effect of self-gravity \citep{2011ApJ...727L..20K,2014ApJ...792...63T}. The log-normal part is narrower in the Ystrong model than the Yweak model. This result is qualitatively consistent with \citet{2011ApJ...730...40P}, who simulated supersonic, self-gravitating, MHD turbulence. In the Ystrong model, the extended tail is much narrower than that in the Yweak model. This is due to the higher magnetic field pressure that suppresses density enhancement of gas in the Ystrong model compared to that in the Yweak model.

\section{Numerical results}\label{results}
We show numerical simulation results of colliding cloud models. Isolated cloud numerical simulation results are used for comparison. Numerical simulation results of isolated cloud models are given in Appendix. 
\subsection{Weak $B_0$ models}\label{weak-section}
\subsubsection{Collision induced structure}
\begin{figure*}
    \begin{center}
    \includegraphics[width=0.7\textwidth]{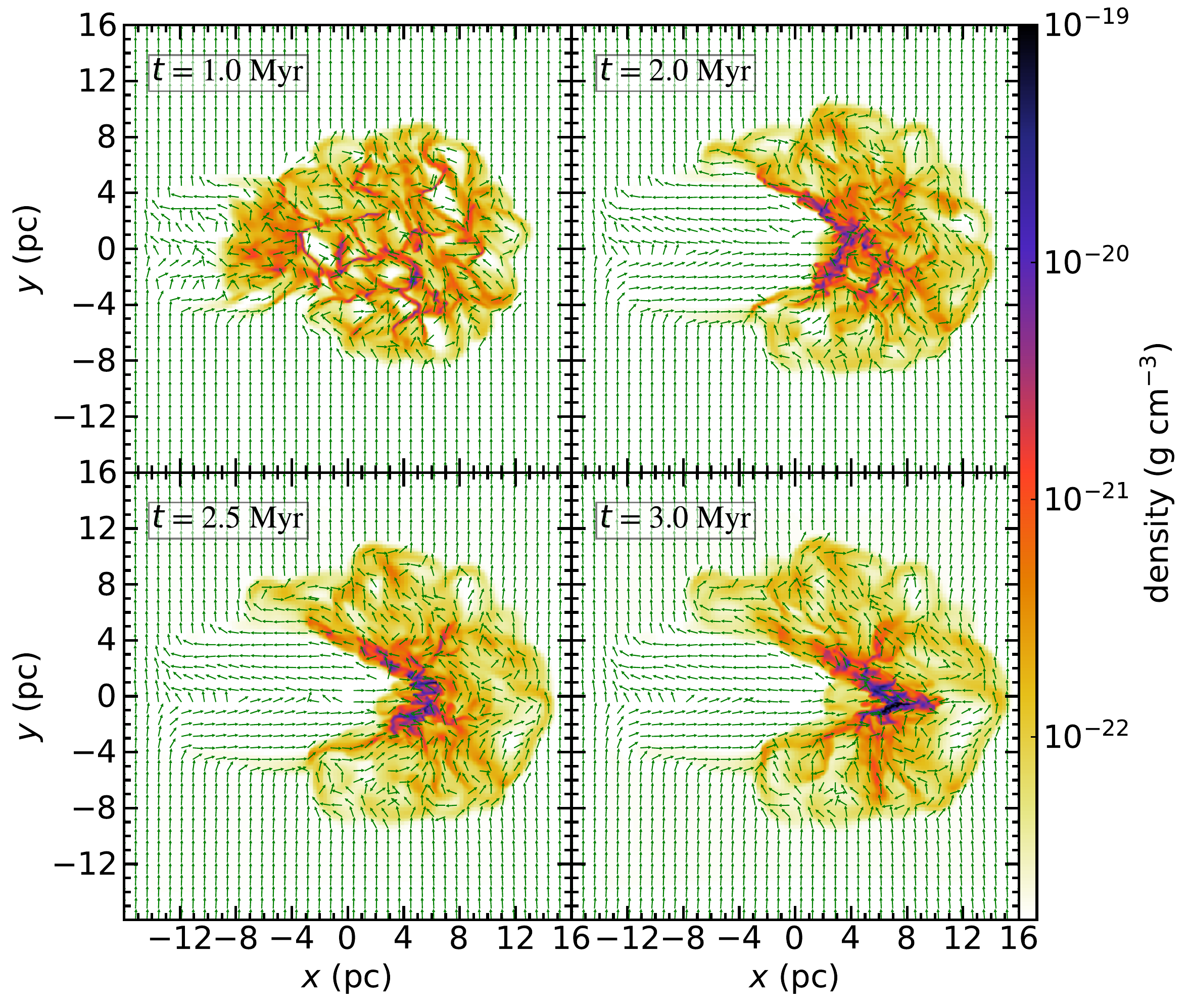}
    \end{center}
    \caption{Slice plots of the gas density in $z$ = 0 pc at $t$ = 1.0 Myr (top left), 2.0 Myr (top right), 2.5 Myr (bottom left), and 3.0 Myr (bottom right) for the Yweak model. The arrows show normalized vectors and the color bar shows gas density, same as in figure \ref{fig:1}.}
    \label{fig:yweak5} 
\end{figure*}
\begin{figure*}
 \begin{tabular}{cc}
 \begin{minipage}[t]{0.5\hsize}
    \begin{center}
    \includegraphics[width=0.75\textwidth]{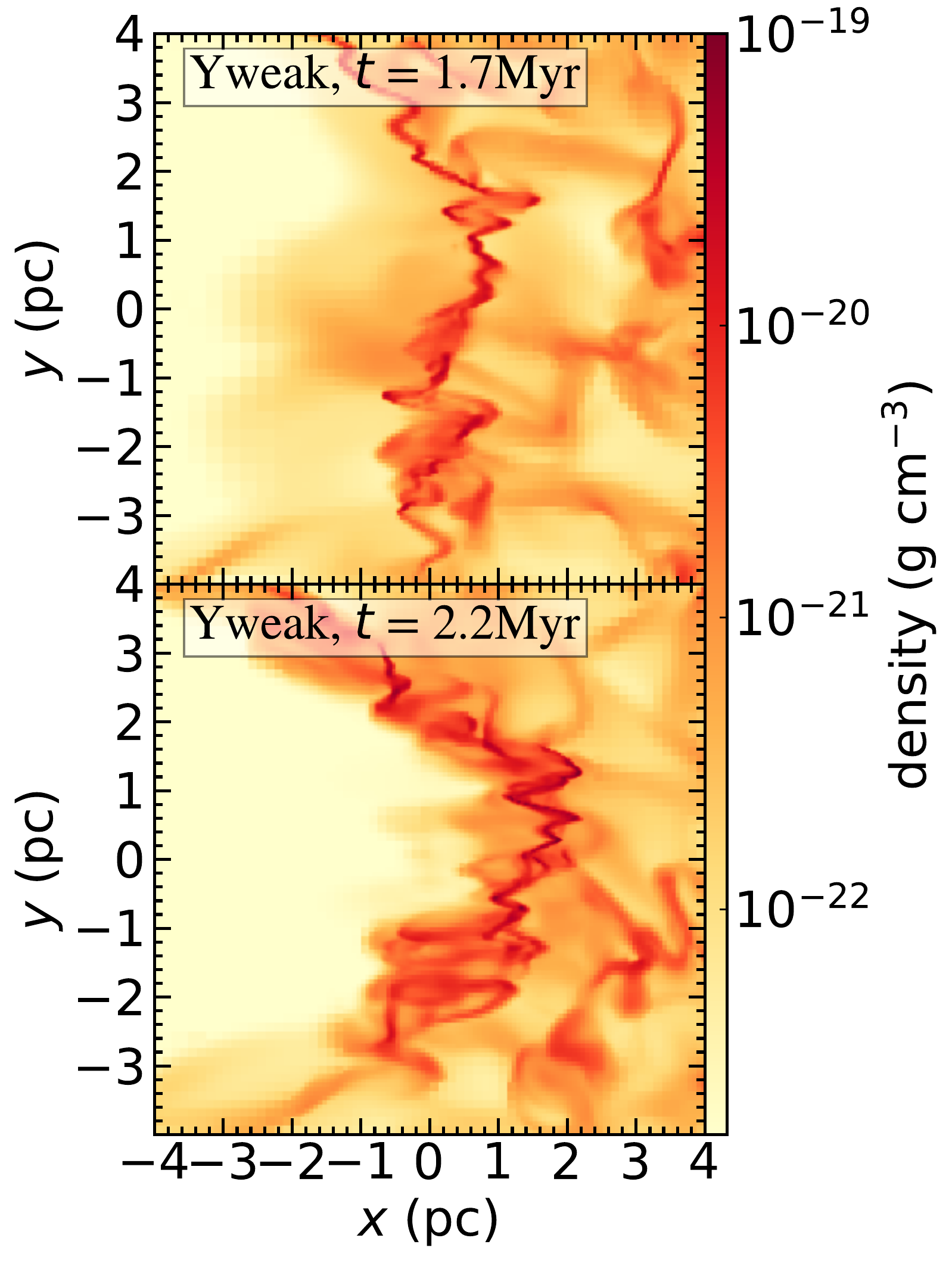}
    \end{center}
      \end{minipage}
    \begin{minipage}[t]{0.5\hsize}
    \begin{center}
    \includegraphics[width=0.7\textwidth]{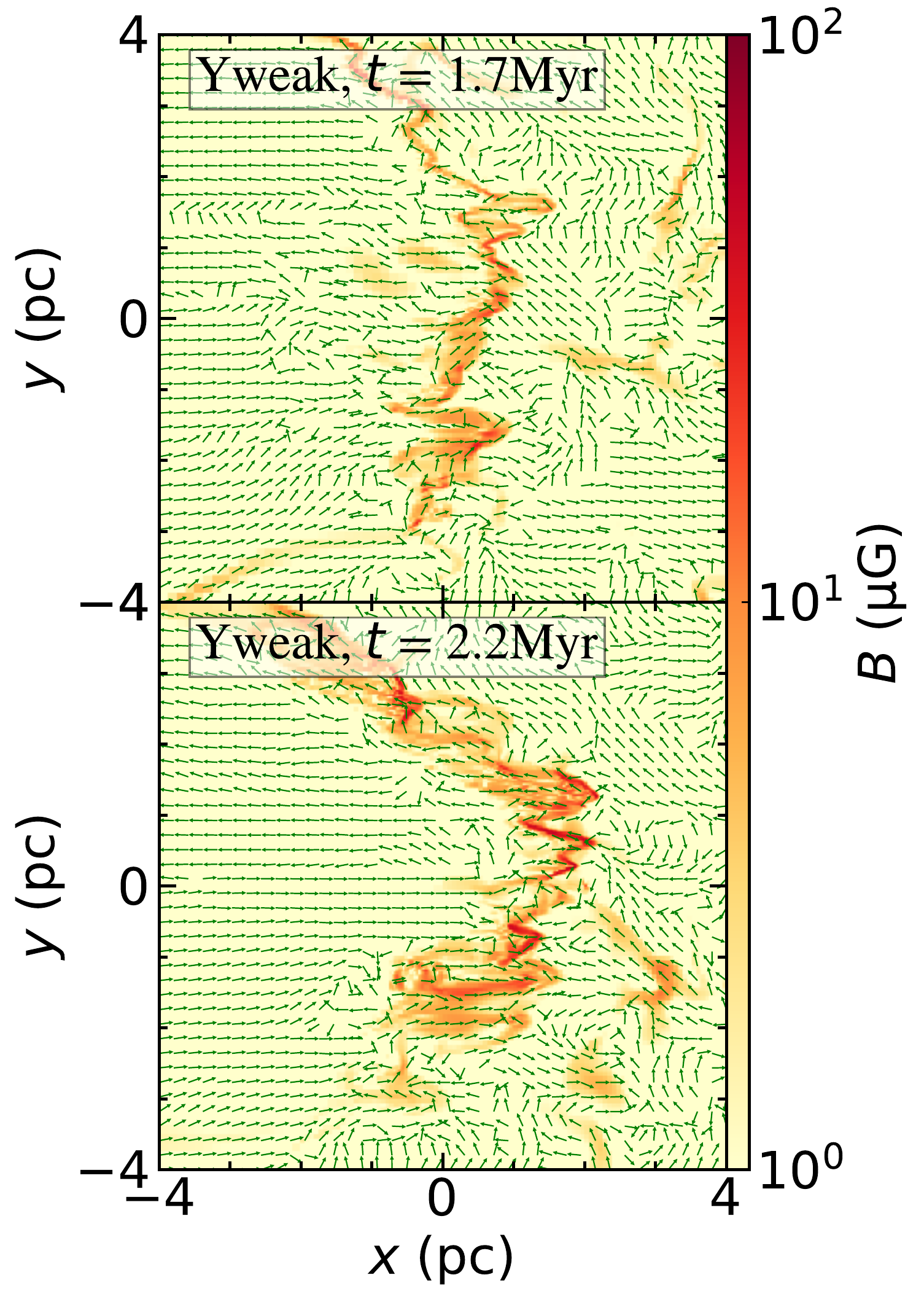}
    \end{center}
    \end{minipage}
     \end{tabular}
       \caption{Slice plots of the gas density (left) and the magnetic field strength (right) for the shocked layer in close-up in $z$ = 0 pc at $t$ = 1.7 Myr (top) and 2.2 Myr (bottom) for the Yweak model. Color bars of the gas density (left) and magnetic field strength (right) are shown on the right edge of each panel. In the right panels, the arrows show normalized vectors, same as in figure \ref{fig:1}. The $x$-coordinates are shifted by 1 pc and 3 pc to the collision velocity direction in the top and bottom panels, respectively.}
        \label{fig:yweakzoom} 
\end{figure*}

As a typical result of the weak $B_0$ models, we show numerical results in of the Yweak model in figure \ref{fig:yweak5}, where $\boldsymbol B_0$ is perpendicular to the collision axis.
As shown in the left-hand panel of figure \ref{fig:2}, the magnetic fields in the clouds are expected to have a minor effect on the gas motion induced by collision, since the Alfv\'{e}n speed in the clouds is much less than the collision speed, 10 km s$^{-1}$. 
In figure \ref{fig:yweak5}, we show the time evolution of the Yweak model at four epochs, $t$ = 1.0 Myr, 2.0 Myr, 2.5 Myr, and 3.0 Myr. 
At $t$ = 1.0 Myr, the two clouds already touch each other, and a thin shocked layer is formed at the interface of the two clouds.
At $t$ = 2.0 Myr, the mass of the shocked layer increases from $t$ = 1.0 Myr, and a cavity in the left-hand side of the shocked layer is formed by the small cloud's penetration into the large cloud. 
The time evolution of colliding clouds in the Yweak model is similar to hydrodynamic simulations of CCC by \citet{2014ApJ...792...63T}, \citet{2018PASJ...70S..58T}, and \citet{2018PASJ...70S..54S}. 

The shocked layer shows quasi-periodic spatial shifts away from a line perpendicular to the collision axis located along the collision interface with scales of less than 1 pc at $t$ = 2.0 Myr. 
Figure \ref{fig:yweakzoom} shows the close-up slice images of gas (left-hand panels) and magnetic field strength (right-hand panels) near the shocked layer at $t$ = 1.7 Myr (upper panels) and 2.2 Myr (lower panels).  
The spatial shifts should be due to the nonlinear thin shell instability (NTSI) \citep{1994ApJ...428..186V,2010MNRAS.405.1431A}.
Figure \ref{fig:yweakzoom} shows that the spatial shifts develop with time between $t$ = 1.7 Myr and $t$ = 2.2 Myr.  
Dense gas concentrations are formed at the extremes of the spatial shifts of the shocked layer.  
The distribution of magnetic field strength is similar to the gas distribution, as shown in figure \ref{fig:yweakzoom}. This implies that the magnetic field plays a minor role in the evolution of the shocked layer. This can be explained as follows. If a magnetic field is strong enough to control the gas flow, the gas flow along the magnetic field is easier than the transverse flow. In this case, the gas distribution can be very different from the magnetic field distribution. If magnetic fields are too weak to affect the gas flow, the gas flow will change the magnetic field structure, and the distribution of magnetic field strength can be similar to the gas distribution for a highly turbulent magnetic field as in the post-shock gas in the clouds.
At $t$ = 2.5 Myr, the shocked layer shape is more concave than $t$ = 2.0 Myr, with additional substructures developing in the shocked layer, as shown in figure \ref{fig:yweak5}. This is due to the shrinking of the central part of the shocked layer along the collision axis caused by converging flow in the shocked layer.
From $t$ = 2.5 Myr to 3.0 Myr, the concave shape develops further with additional substructures developing in the shocked layer and dense mass concentrations are formed near $x$ $\sim$ 5 pc and $y$ $\sim$ 0 pc in figure \ref{fig:yweak5}, which is the bottom of the concave shape. 
A similar evolution is found in the numerical results of the other models with the weak $B_0$, implying that the magnetic field plays only a minor role in the evolution of the shocked layer in the weak $B_0$ models.

\begin{figure*}
     \begin{center}
     \includegraphics[width=1.\textwidth]{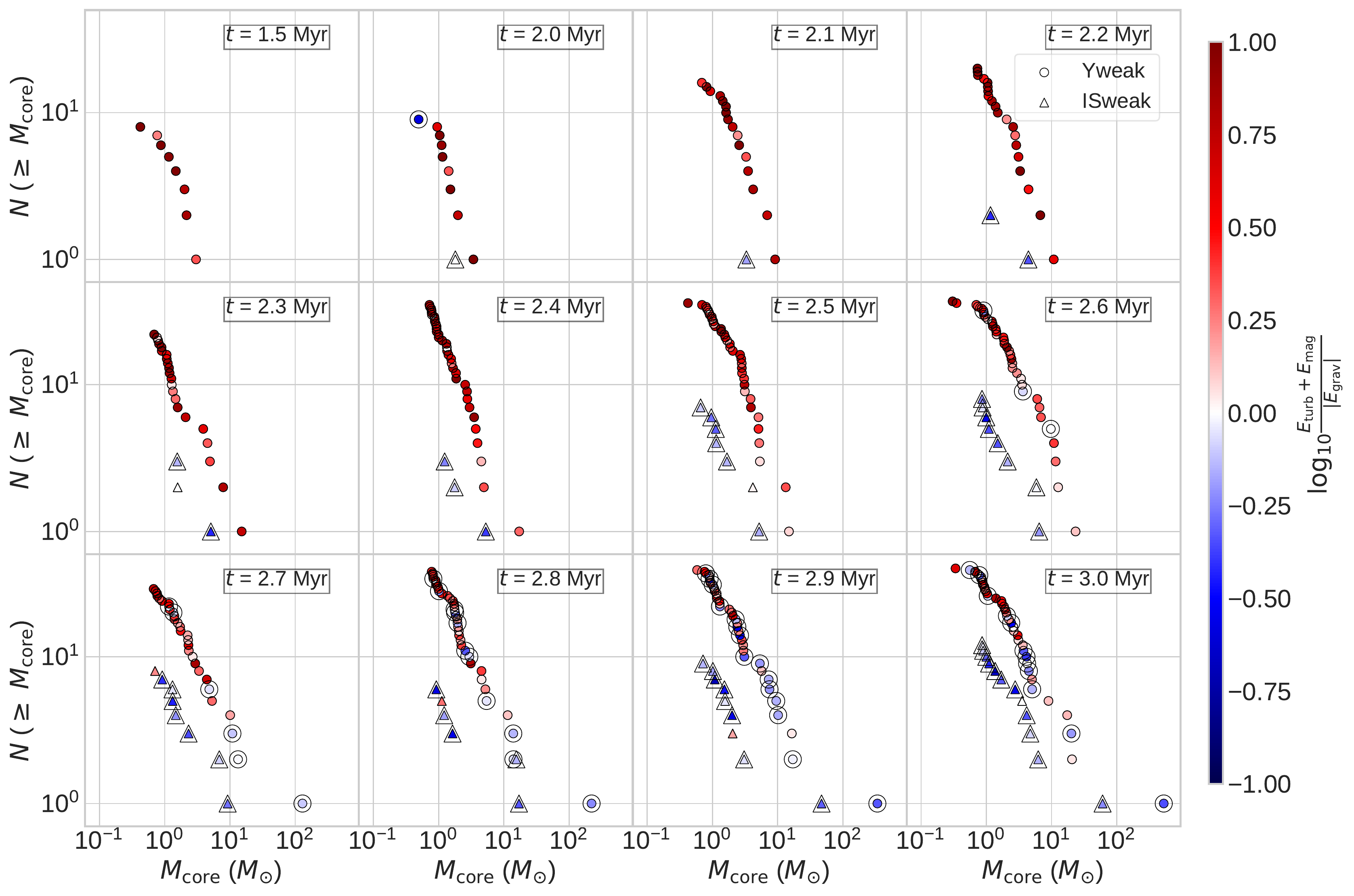}
     \end{center}
     \caption{Cumulative core mass distributions shown by filled circles and filled triangles at $t$ = 1.5 Myr and $t$ = 2.0 Myr to 3.0 Myr for the Yweak and ISweak models, respectively. The color bar in the right-hand side shows the energy ratio of turbulent energy plus magnetic field energy to self-gravitational energy (absolute value). The gravitationally bound cores are marked by larger open circles and larger open triangles for the Yweak and ISweak models, respectively.}
    \label{fig:yWeakCMF} 
\end{figure*}
\begin{figure*}
    \begin{center}
    \includegraphics[width=1.\textwidth]{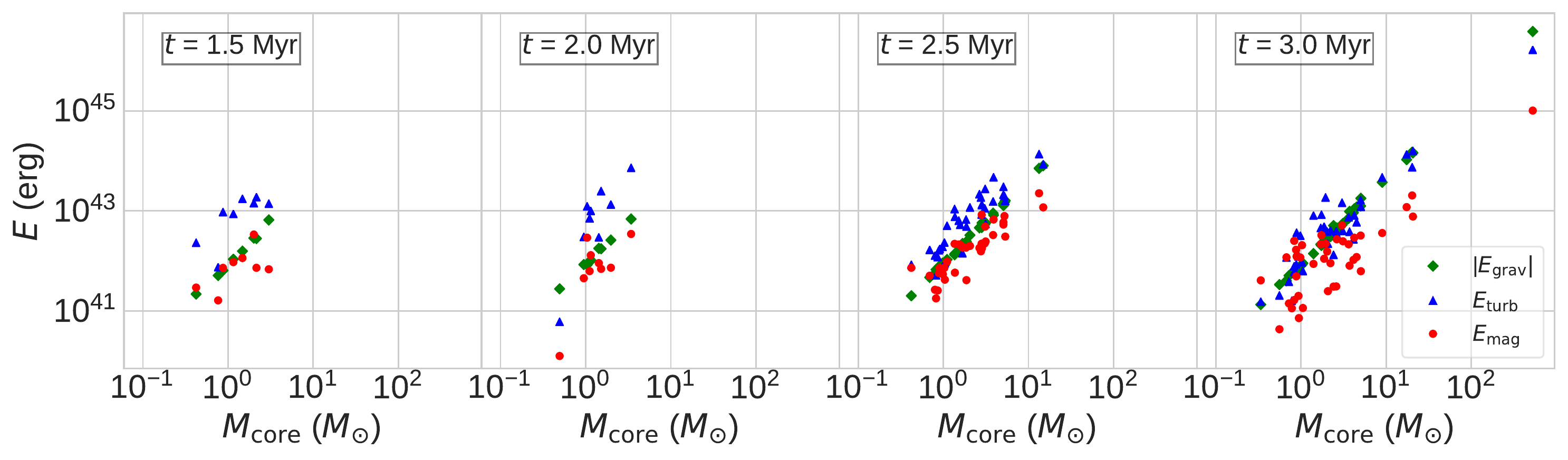}
    \end{center}
    \caption{Energies of the self-gravity (absolute value) $|E_{\rm grav}|$ (diamond), turbulence $E_{\rm turb}$ (triangle), and magnetic field $E_{\rm mag}$ (circle) of the dense cores at $t$ = 1.5 Myr (left), 2.0 Myr (middle left), 2.5 Myr (middle left), and 3.0 Myr (right) for the Yweak model.}
    \label{fig:yWeakEnergy} 
\end{figure*}

\subsubsection{Dense core formation and evolution}\label{dense-core}
In order to study dense core formation and evolution, we define a dense core by the threshold density, $\rho_{\rm th}$ = 5 $\times$ 10$^{-20}$ g cm$^{-3}$, which is in the range of the typical density of molecular cores \citep{2007ARA&A..45..339B}. 
We define dense cores by following steps: (1) selection of cells with $\rho$ $\ge$ $\rho_{\rm th}$ as dense cells, (2) grouping together with neighboring dense cells, and (3) rejection of those groups with cell number less than 27.
This minimum cell number condition is used to obtain a good resolution of dense cores.

We show a cumulative dense core mass distribution, $N$ ($\ge$ $M_{\rm core}$), which is number of cores with mass more than $M_{\rm core}$, at $t$ = 1.5 Myr and 2.0 Myr to 3.0 Myr in figure \ref{fig:yWeakCMF}. We also show the cumulative dense core mass distribution of the isolated cloud model with the weak magnetic field using filled triangles in figure \ref{fig:yWeakCMF}. 
Figure \ref{fig:yWeakCMF} clearly shows that more massive dense cores are formed in the colliding clouds model than the isolated model. 
The dense cores are already formed at $t$ = 1.5 Myr and the core number rapidly increases from $t$ = 2.0 Myr to $t$ = 2.5 Myr in the colliding clouds. 
In the isolated cloud, the first dense core forms at a later epoch than in the colliding clouds, and the total number of dense cores is also much smaller compared to the colliding clouds at each epoch shown.
At $t$ = 3.0 Myr, four massive dense cores with their masses greater than 10 $M_{\odot}$ are formed in the colliding clouds. 

We check the gravitational stability of each dense core by comparing its turbulent energy, $E_{\rm turb}$, its magnetic field energy, $E_{\rm mag}$, and an absolute value of its self-gravitational energy, $|E_{\rm grav}|$. Here we ignore its thermal energy, since temperature of the dense core is nearly 10 K and the sound speed of gas at this temperature is much smaller than the turbulent velocity and the Alfv\'en speed. 
$E_{\rm turb}$ is given for a dense core by
\begin{equation}
E_{\rm turb}=\sum_i \frac{1}{2}m_i|\boldsymbol v_i-\boldsymbol v_{\rm mean}|^2,
\end{equation}
where $i$ is an index of a dense cell in the dense core, the sum is made over all cells belonging to the dense core, $m_i$ is the mass of the dense cell $i$, and $\boldsymbol v_{\rm mean}$ is the mean velocity of the dense core given by
\begin{equation}
\boldsymbol v_{\rm mean}= \frac{\sum_im_i\boldsymbol v_i}{\sum_im_i}.
\end{equation}
$E_{\rm mag}$ is given by 
\begin{equation}
E_{\rm mag}=\sum_i \frac{\boldsymbol B_i \cdot \boldsymbol B_i}{8\pi }V_i,
\end{equation}
where $\boldsymbol B_i$ and $V_i$ are the magnetic field flux density vector and volume, respectively, of the dense cell $i$.
We estimate  $|E_{\rm grav}|$ by
\begin{equation}
|E_{\rm grav}| = \frac{3GM_{\rm core}^2}{5\ensuremath{\langle R \rangle}},
\end{equation}
where $G$ is the gravitational constant, $M_{\rm core}$ is the mass of the dense core, and $\ensuremath{\langle R \rangle}$ is given by 
\begin{equation}
\ensuremath{\langle R \rangle} = {\left(\frac{3V_{\rm core}}{4\pi}\right)^{1/3}},
\end{equation}
where $V_{\rm core}$ is total volume of the dense core. 
In figure \ref{fig:yWeakEnergy}, we show these energies.
If $(E_{\rm turb}+E_{\rm mag})$ $\leq$ $|E_{\rm grav}|$, we can expect that the dense core is gravitationally bound, and we call such a dense core a gravitationally bound core. 
Since a dense core is defined by the condition of $\rho$ $\geq$ $\rho_{\rm th}$, its free-fall time $t_{\rm ff}$ $\leq$ 0.3 Myr. 
Many cores are gravitationally bound at $t$ = 3.0 Myr, as shown in figure \ref{fig:yWeakCMF}. 
The main reason for the formation of gravitationally bound cores is the turbulent energy dissipation, as shown in figure \ref{fig:yWeakEnergy}. 
Figure \ref{fig:yWeakEnergy} also shows that the turbulent energy is still comparable to the gravitational binding energy in the bound cores at $t$ = 3.0 Myr. In the monolithic collapse scenario by \citet{2003ApJ...585..850M}, a primary star formed in a massive bound core with such large turbulence will be massive. If we apply the monolithic collapse scenario to those massive bound cores, we can expect massive star formation in them. The most massive dense core is formed in the dense gas region near the bottom of the concave structure of the shocked layer. This core becomes gravitationally bound at $t$ = 2.7 Myr, and its mass is 127 $M_{\odot}$ at this epoch.
Rapid mass increase of this core from $t$ = 2.7 Myr to 3.0 Myr is due to gas accretion on the core, since this core is in the gas dense region near the bottom of the concave structure of the shocked layer. In the isolated cloud with the weak $B_0$, the first bound core forms at $t$ = 2.0 Myr and its mass is $\sim$ 2 $M_{\odot}$.  
The mass of the bound core increases to $\sim$ 5 $M_{\odot}$ after its free-fall time. 
The other bound cores are formed with mass $\sim$ 1 $M_{\odot}$, and the mass evolution of these cores is similar to the first bound one. In the monolithic collapse scenario, a star will form with the free-fall timescale of the bound core. In these bound cores, we can expect low- or intermediate-mass star formation according to the monolithic scenario, since the core masses are less than 10 $M_{\odot}$ after their free-fall time from their formation epoch of gravitationally bound cores.
\subsection{Strong $B_0$ models}\label{strong}
\subsubsection{Collision induced structure}
\begin{figure*}
    \begin{center}
    \includegraphics[width=0.7\textwidth]{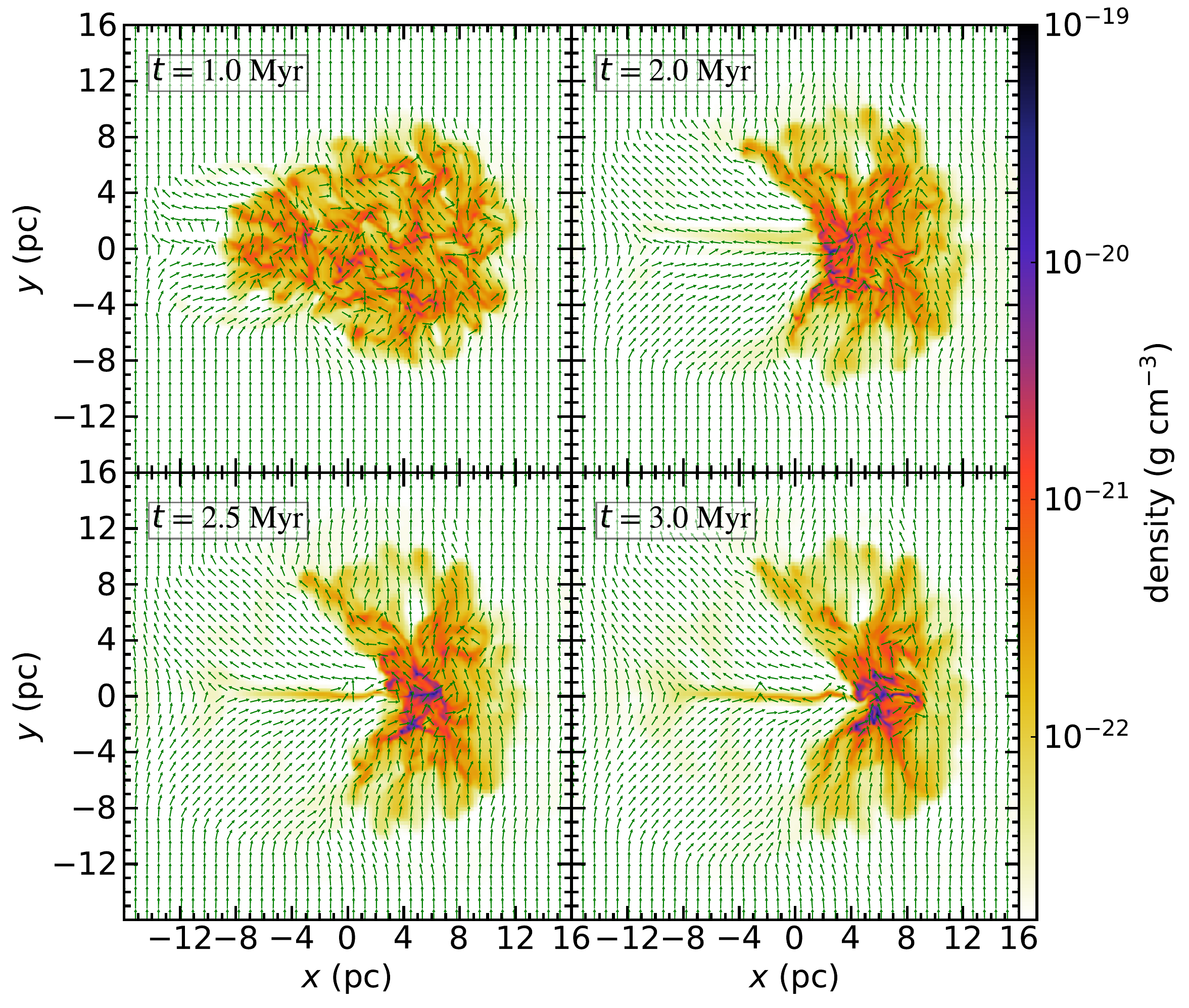}\\
    \end{center}
    \caption{Same as figure \ref{fig:yweak5}, but for the Ystrong model.}
    \label{fig:ystrong5} 
\end{figure*}

As a typical result of the strong $B_0$ models, we show numerical results of the Ystrong model in which $B_0$ = 4.0 $\mu$G and its direction is perpendicular to the collision axis in figure \ref{fig:ystrong5}. 
In this model, the collision speed, 10 km s$^{-1}$, is larger than the Alfv\'{e}n speed of gas in both clouds, as shown in the right-hand panel of figure \ref{fig:2}. 
The formation of the shocked layer and formation of the cavity are similar to the weak $B_0$ models, as shown in subsection \ref{weak-section}. However, the cavity produced by the collision of the small cloud displays a wider opening angle than in the Yweak model, as shown in figure \ref{fig:yweak5} and figure \ref{fig:ystrong5}. This can be explained by the smaller Alfv\'{e}n Mach number in terms of collision speed ($v_{\rm coll}$), $M_{\rm A}$ = $v_{\rm coll}/v_{\rm A}$, in this model than in the Yweak model (see figure \ref{fig:2}), as in the MHD bow shock wave formed in the solar wind around a planet \citep{1984JGR....89.2708S}. The shocked layer produced by the CCC is much thicker than that seen in the Yweak model.
This may be due to smaller $M_A$ and the larger magnetic pressure in the shocked layer in the Ystrong model than the Yweak model. 

 \begin{figure*}[ht]
  \begin{tabular}{cc}
 \begin{minipage}{0.45\hsize}
    \begin{center}
    \includegraphics[width=0.98\textwidth]{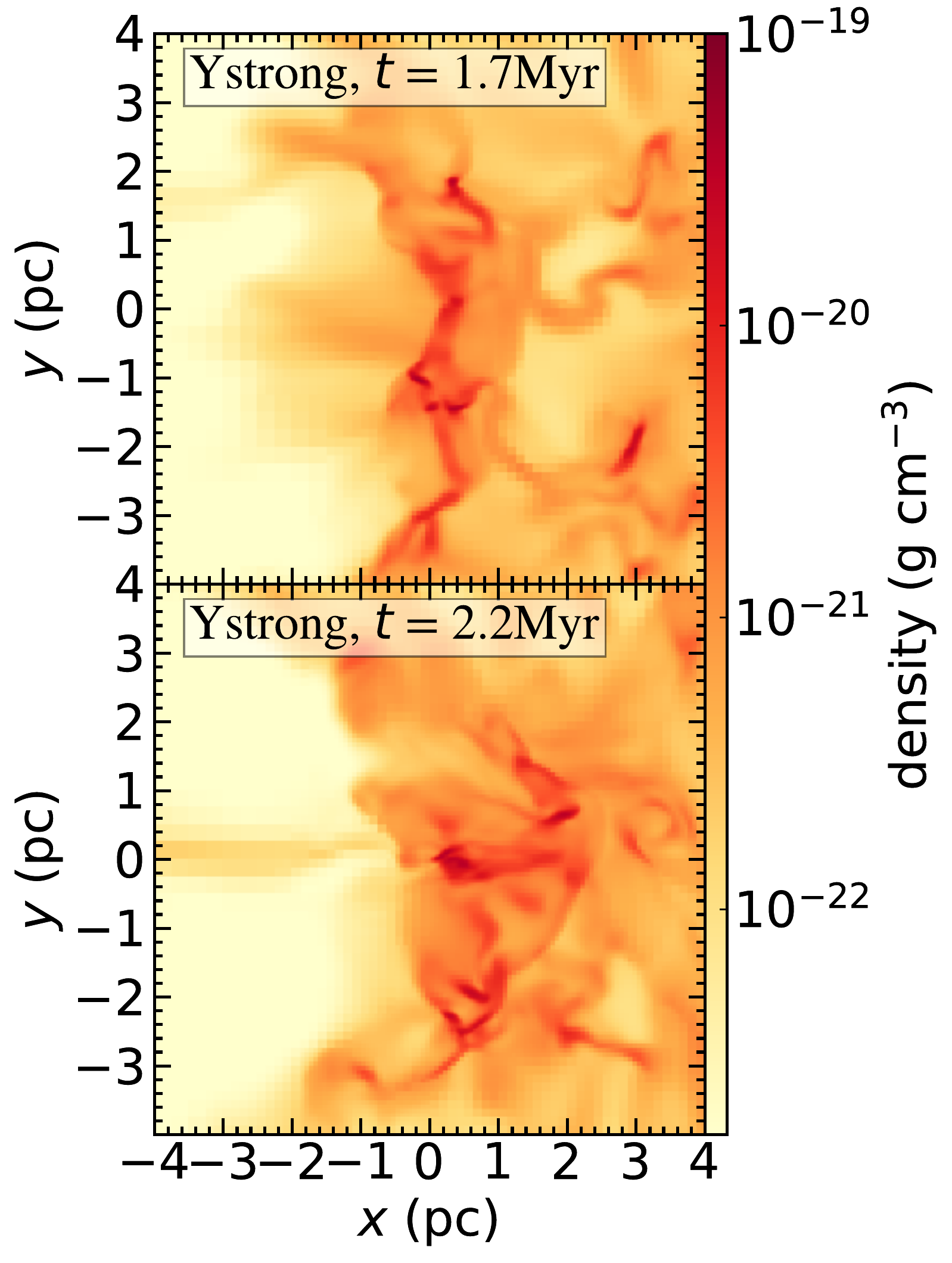}
     {\\(a)  density}
    \end{center}
      \end{minipage}
    \begin{minipage}{0.45\hsize}
    \begin{center}
    \includegraphics[width=0.9\textwidth]{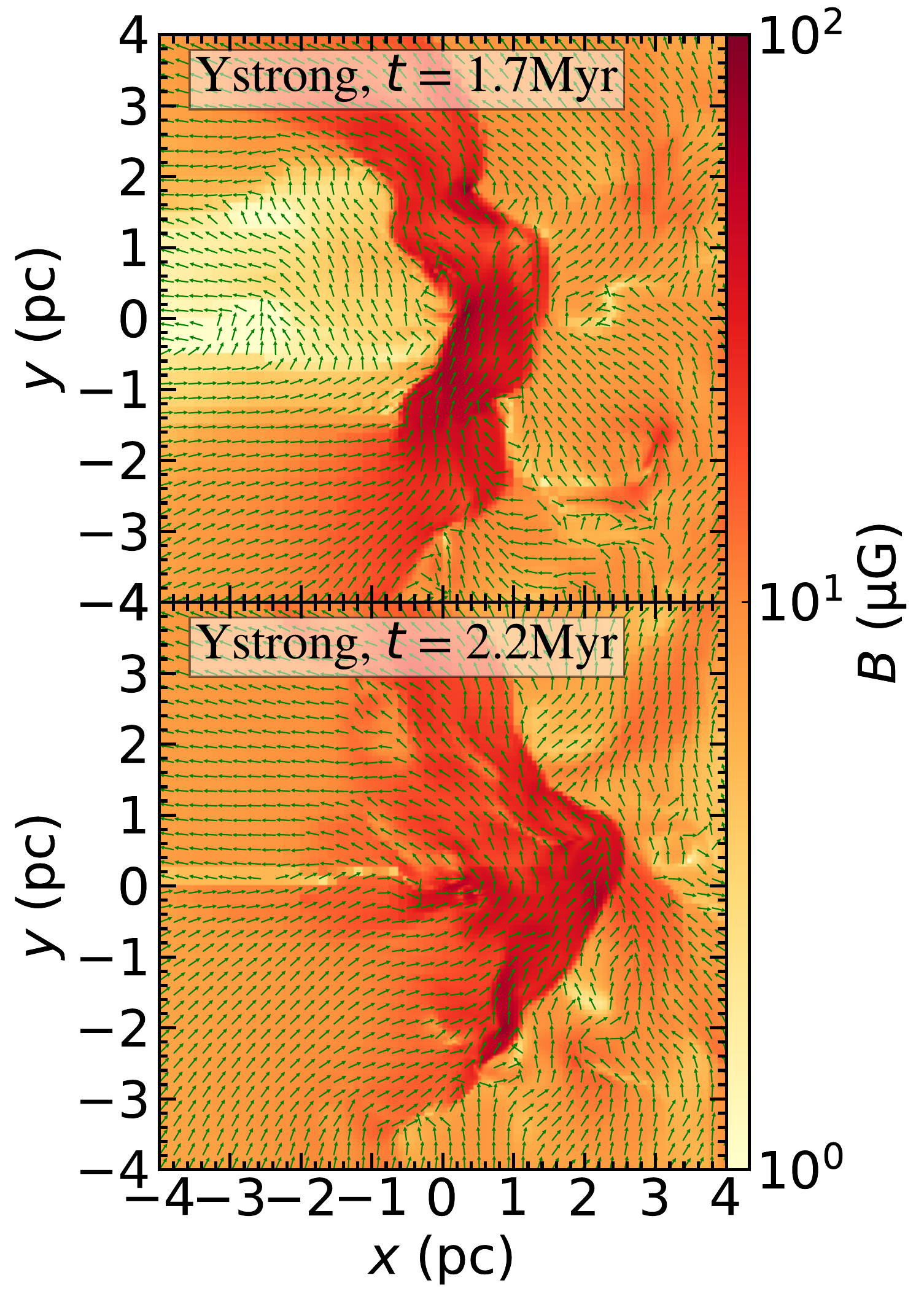}
    {\\(b) magnetic field strength}
    \end{center}
    \end{minipage}
     \end{tabular}\\ \\
       \caption{Same as figure \ref{fig:yweakzoom}, but for the Ystrong model.}
       \label{fig:ystrongzoom} 
\end{figure*}

Figure \ref{fig:ystrongzoom} is a close-up slice image of the shocked layer at $t$ = 1.7 and 2.2 Myr. This figure shows that there are no clear quasi-periodic spatial shifts of the layer, contrary to the Yweak model. The strong magnetic fields suppress the NTSI in the Ystrong model. 
The NTSI in an MHD flow was studied by \citet{2007ApJ...665..445H}. They show that the magnetic fields weaken the NTSI. 
Density fluctuations of larger sizes than the Yweak model are formed in the shocked layer, as shown in figures \ref{fig:yweakzoom} and \ref{fig:ystrongzoom}. The density fluctuations of larger sizes can be formed by Richtmyer-Meshkov instability, by which density fluctuations in the pre-shock region are enhanced in the post-shock gas, as shown by \citet{Inoue_2009} and \citet{Mizuno_2010}. For example, \citet{2012ApJ...759...35I} simulated a colliding flow with inhomogeneities of gas density and showed that the post-shock gas is highly turbulent and that density fluctuations in the post-shock gas are created by the development of the Richtmyer-Meshkov instability. More detailed analysis of the Richtmyer-Meshkov instability in our models is beyond the scope of this paper.

The density and magnetic field enhancements are not coincident, as shown in figure \ref{fig:ystrongzoom}. Distribution of the magnetic field strength is much smoother than the density enhancements.
This indicates that the turbulent magnetic fields on a large scale are enhanced by the CCC flow, and gas moves along the smaller scale magnetic fields to create further density enhancements in the shocked layer. 
In this way, a difference in density and magnetic field enhancements is produced.

\subsubsection{Dense core formation and evolution}

\begin{figure*}
    \begin{center}
    \includegraphics[width=1.0\textwidth]{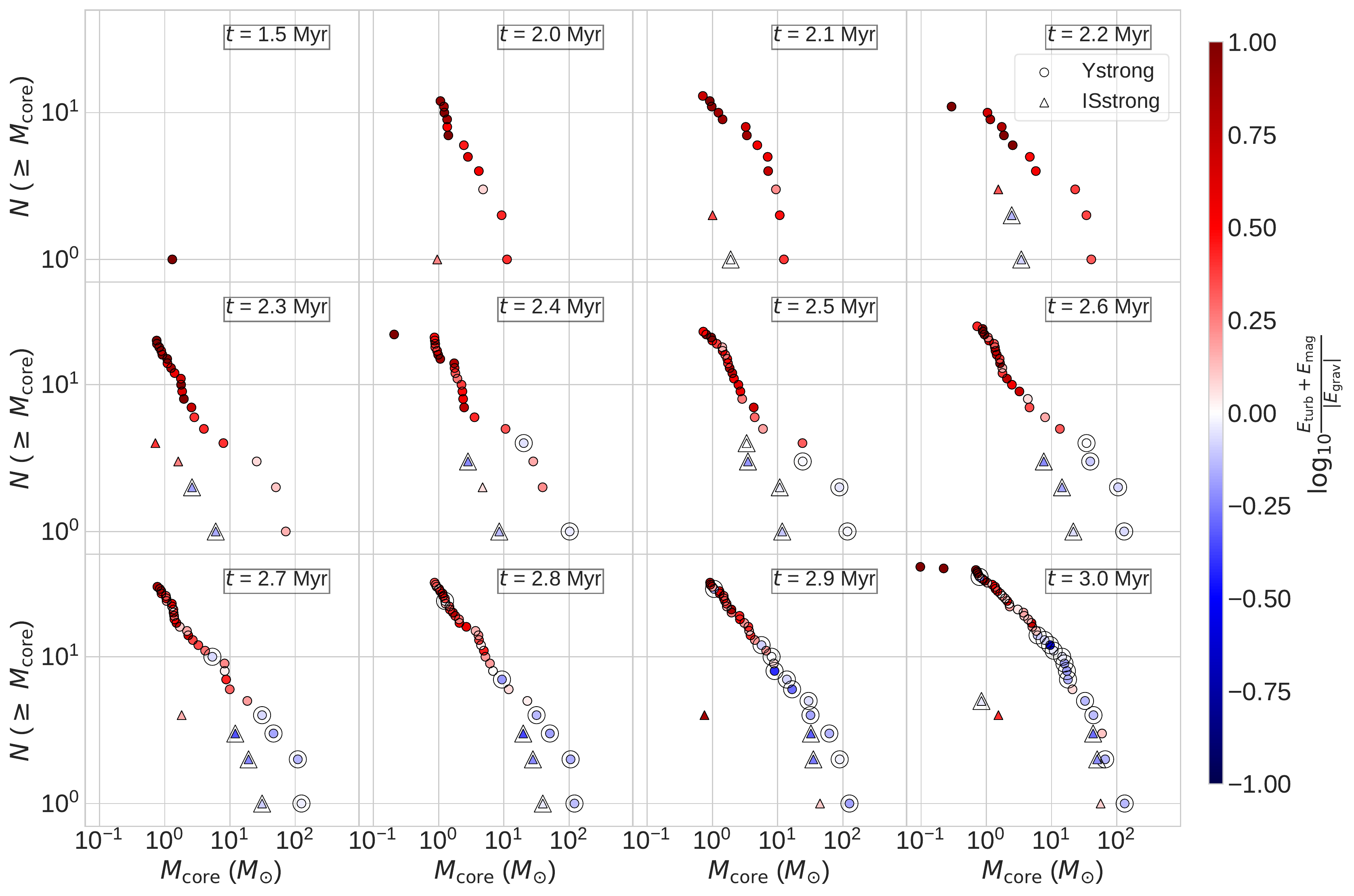}
    \end{center}
    \caption{Same as figure \ref{fig:yWeakCMF}, but for the Ystrong and ISstrong models.}
    \label{fig:cmf-strongy} 
\end{figure*}
\begin{figure*}
    \begin{center}
    \includegraphics[width=1.0\textwidth]{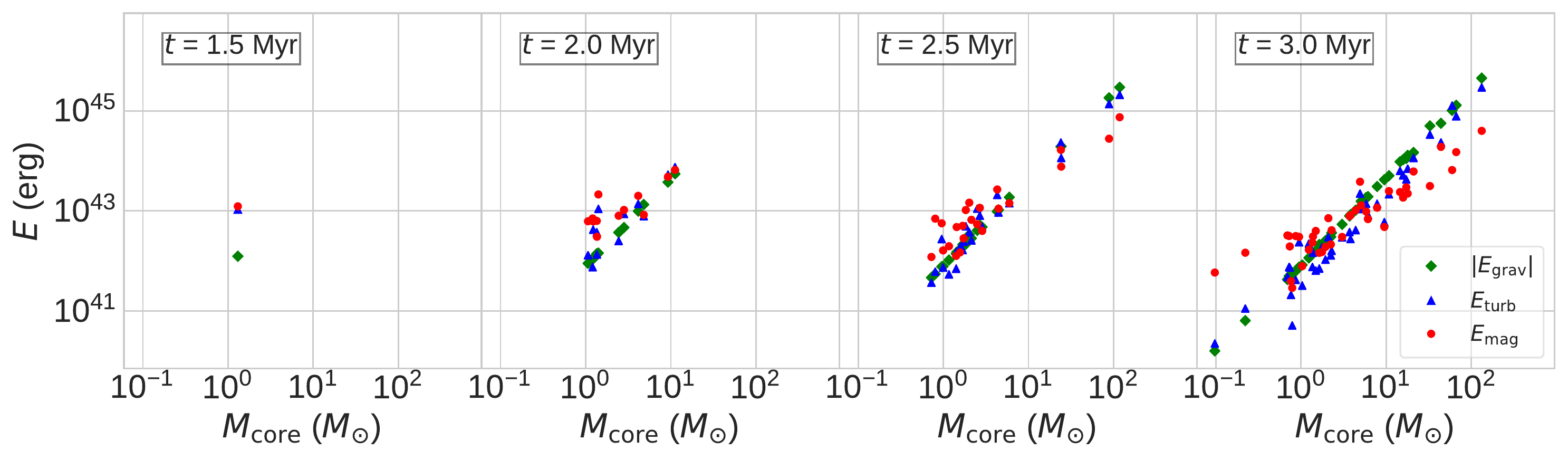}
    \end{center}
    \caption{Same as figure \ref{fig:yWeakEnergy}, but for the Ystrong model.}
    \label{ver3thesisimage10} 
\end{figure*}

We show the time evolution of the cumulative core mass distribution in the Ystrong model in figure \ref{fig:cmf-strongy}. The total number of dense cores formed in the Ystrong model is less than that in the Yweak model (see figure \ref{fig:yWeakCMF}) during the early phase of collision ($t$ = 1.5 Myr). This is due to the suppression of the NTSI in the smaller scales in the Ystrong model. More massive dense cores are formed in the Ystrong model than the Yweak model at $t$ $\gtrsim$ 2.0 Myr, and the number of the massive dense cores with masses more than 10 $M_{\odot}$ is also greater than the Yweak model at $t$ = 3.0 Myr. 
The massive dense core formation is due to gas accumulation to massive dense cores in the dense gas regions in the thick shocked layer, as shown in figure \ref{fig:ystrongzoom}. 

The time evolution of the magnetic field energy, the turbulent energy, and the absolute value of self-gravitational energy of each dense core with its mass is shown in figure \ref{ver3thesisimage10}
from $t$ = 1.5 Myr to $t$ = 3.0 Myr. 
At $t$ = 1.5 Myr and 2.0 Myr, the magnetic field energies are larger than the absolute self-gravitational energies in all cores.
However, at $t$ = 2.5 Myr and 3.0 Myr, the absolute self-gravitational energies are dominant in the massive dense cores. This change is mainly due to their mass increase by gas accumulation.

\begin{figure*}
    \begin{center}
    \includegraphics[width=0.7\textwidth]{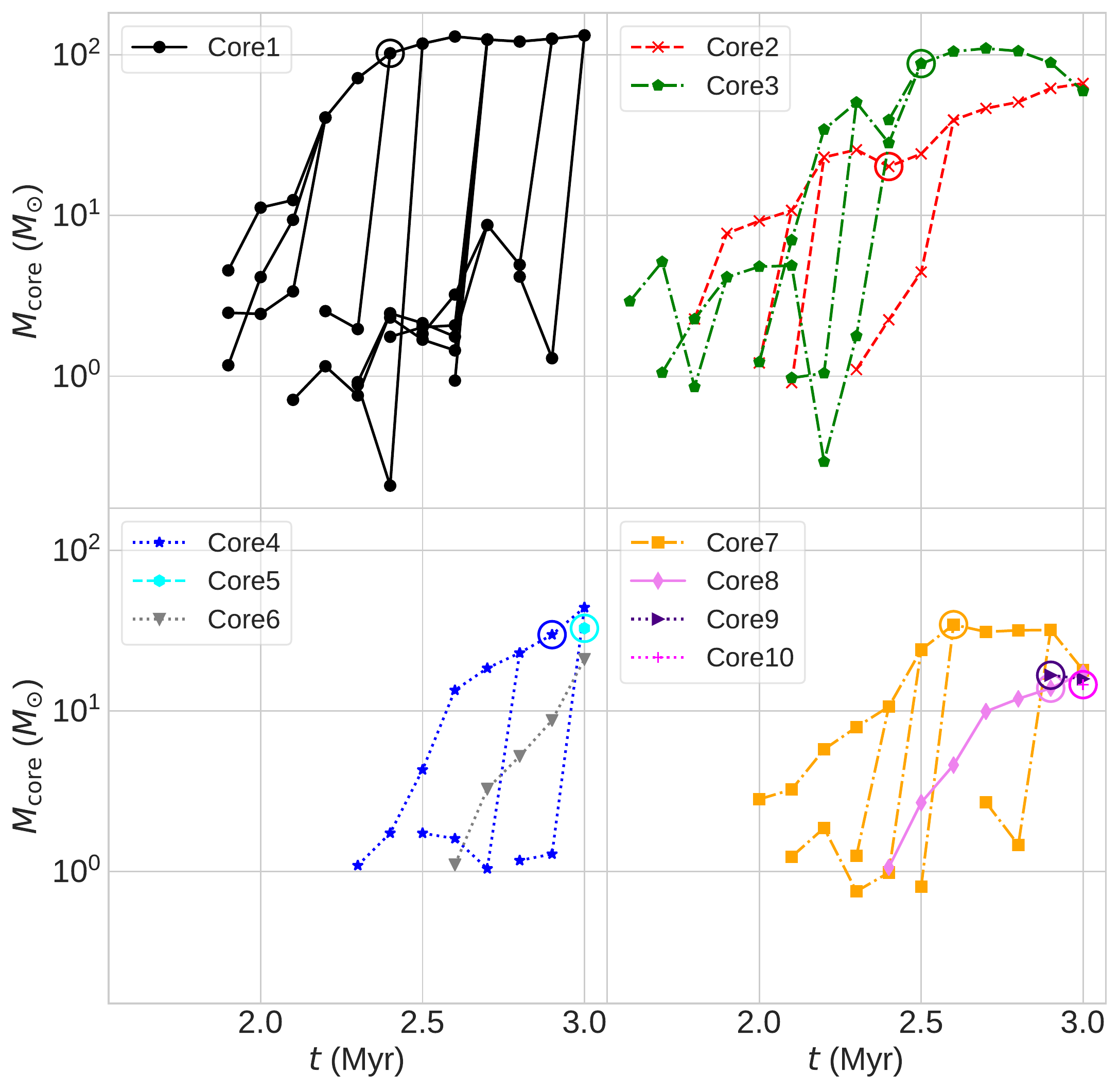}
    \end{center}
    \caption{Mass evolution of top ten massive dense cores at $t$ = 3.0 Myr in the Ystrong model. The cores are numbered in descending order of their mass. Merging events are shown by joining of the evolutionary tracks of the core. Open circles show the formation epoch of gravitationally bound cores.}
    \label{ver3thesisimage14} 
\end{figure*}

The mass evolution of the top 10 massive dense cores and the epoch for them to become gravitationally bound is shown in figure \ref{ver3thesisimage14}.
In this figure, we also show merger trees that indicate dense core mergers. The mass growth of any given dense core is a combination of the accretion of surrounding gas and mergers with other dense cores. If the mass contribution by mergers is not enough to explain the mass growth of the dense core, the gas accretion to the dense core should be a dominant process for the mass growth. We estimate mass contribution of core mergers to the mass growth of the top ten massive dense cores. Figure \ref{ver3thesisimage14} shows that Core1 grows from 5 $M_{\odot}$ at $t$ = 1.9 Myr to 132 $M_{\odot}$ at $t$ = 3.0 Myr with mass contribution of 26 $M_{\odot}$ (20 \%) by mergers of dense cores during this time interval. Core2 grows from 2 $M_{\odot}$ at $t$ = 1.8 Myr to 66 $M_{\odot}$ at $t$ = 3.0 Myr with mass contribution of 7 $M_{\odot}$ (10 \%) by mergers of dense cores during this time interval. Core3 grows from 39 $M_{\odot}$ at $t$ = 2.4 Myr to its peak of 110 $M_{\odot}$ at $t$ = 2.7 Myr with a mass contribution of 28 $M_{\odot}$ (40 \%) by one merger at $t$ = 2.4 Myr. For Core4, mergers contribute 5 \% mass to the core mass growth. For Core7, mergers contribute 10 \% mass up to its peak mass at $t$ = 2.6 Myr. There are no mergers for the rest of the massive dense cores. Merger events play more of a secondary effect in increasing the mass of the top ten massive dense cores, with mass growth primarily a smoother function of time implying accretion dominated, evolution.
Since the free-fall time of these massive dense cores is less than 0.3 Myr, we expect that protostars form in these cores.

We highlight the gravitationally bound cores using larger open circles in figure \ref{fig:cmf-strongy}.
As shown in figure \ref{fig:cmf-strongy}, three massive dense cores become gravitationally bound at $t$ = 2.5 Myr, although these are not gravitationally bound at $t$ = 2.0 Myr. Their masses are larger than 10 $M_{\odot}$. 
At $t$ = 3.0 Myr, most dense cores (nine out of the 10) with more than 10 $M_{\odot}$ become gravitationally bound.
The number of bound cores in the Ystrong model is much larger than the Yweak model in which only two bound cores with a masses more than 10 $M_{\odot}$ are found at $t$ = 3.0 Myr, as shown in figure \ref{fig:yWeakCMF}.
This may be due to the thick shocked layer caused by the strong magnetic field in the Ystrong model. 
The NTSI grows faster for a small-scale shift of the shocked layer. In the Yweak model, the NTSI develops in the small scale shift and results in the accumulation of low-mass gas at the extremes of the quasi-periodic shifts, as shown in figure \ref{fig:yweakzoom}. These gas concentrations move with large irregular velocities, which may suppress further gravitational gas accumulation to the gas concentrations. As a result, this mass growth by gas accretion can be suppressed. 
In the Ystrong model, since the shocked layer is thickened by the strong magnetic fields, as shown in figure \ref{fig:ystrongzoom}, and the dense cores have small irregular velocities by suppression of the NTSI, the dense cores can acquire mass by accretion in the thick shocked layer.
The cumulative core mass distribution in the isolated cloud model with the strong magnetic field, $B_0$ = 4 $\mu$G, is shown for comparison in figure \ref{fig:cmf-strongy} using filled triangle symbols. The bound cores form earlier than the Ystrong model, and the mass of bound cores just formed is less than 3 $M_{\odot}$.
We can expect intermediate-mass star formation in these cores in the free-fall time $\sim$ 0.3 Myr, assuming that the gas mass in these cores accretes to form a single star.
This is very different from colliding clouds models, which hosted bound cores formed with masses greater than 10 $M_{\odot}$. The resulting formation of protostars can be studied using sink particle models, though this is beyond the scope of the work presented here \citep{2010ApJ...713..269F,2018PASJ...70S..54S}.
\subsection{Core mass distribution and gravitationally bound cores}\label{CMF}

\begin{figure*}
    \begin{center}
    \includegraphics[width=0.9\textwidth]{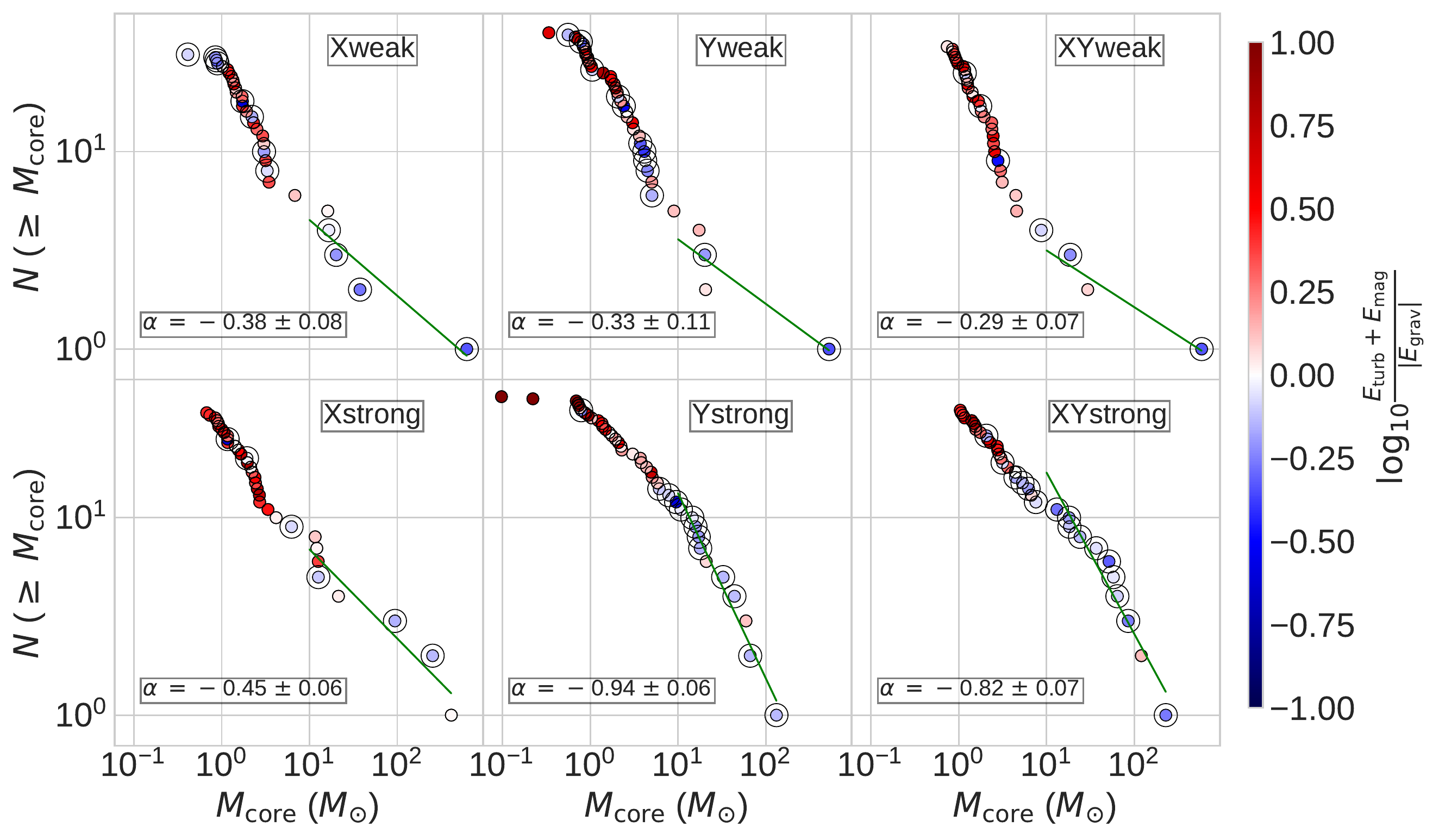}
    \end{center}
    \caption{Cumulative core mass distributions at $t$ = 3.0 Myr for the Xweak (top left panel), Yweak (top middle panel), XYweak (top right panel), Xstrong (bottom left panel), Ystrong (bottom middle panel), and XYstrong (bottom right panel) models. The larger open circles show the gravitationally bound cores. The least square fits with standard deviations for cores with masses greater than 10 $M_{\odot}$ done using equation \ref{cmf} are shown.}
    \label{fig:cmf-all} 
\end{figure*}

We show the core mass functions of all models at $t$ = 3.0 Myr in figure \ref{fig:cmf-all}.
We find many gravitationally bound cores in the strong $B_0$ models at $t$ = 3.0 Myr. 
The core mass distributions of the weak $B_0$ models are very similar to each other. 
In each weak $B_0$ model, there is one exceptionally massive dense core.
In the strong $B_0$ models, the number of dense cores with mass more than 10 $M_{\odot}$ is larger than the weak $B_0$ models.
This indicates that the strong $B_0$ contributes to the formation of a greater number of massive dense cores. 
The number of massive dense cores in the Xstrong model is slightly smaller than the Ystrong and XYstrong models, yet it is still greater than those in the weak $B_0$ models (Xweak, Yweak, and XYweak models). 
These results indicate that the strong magnetic field parallel to the collision axis (as in the Xstrong) is less effective in suppressing the NTSI and in keeping the shocked layer thick,
compared with the strong magnetic field with orientations oblique or perpendicular to the collision axis (as in the Ystrong and XYstrong models).

The mass of the most massive dense cores attains more than 100 $M_{\odot}$ in the last 1 Myr in all models, since we cannot find such massive dense cores at $t$ = 2.0 Myr. Since the free-fall time of these cores is less than 0.3 Myr, we can expect rapid protostar formation before $t$ = 3.0 Myr. If massive stars are formed in those colliding clouds, we can expect very strong feedback from these massive stars, as shown by \citet{2018PASJ...70S..54S}. 

\begin{figure*}
    \begin{center}
    \includegraphics[width=0.7\textwidth]{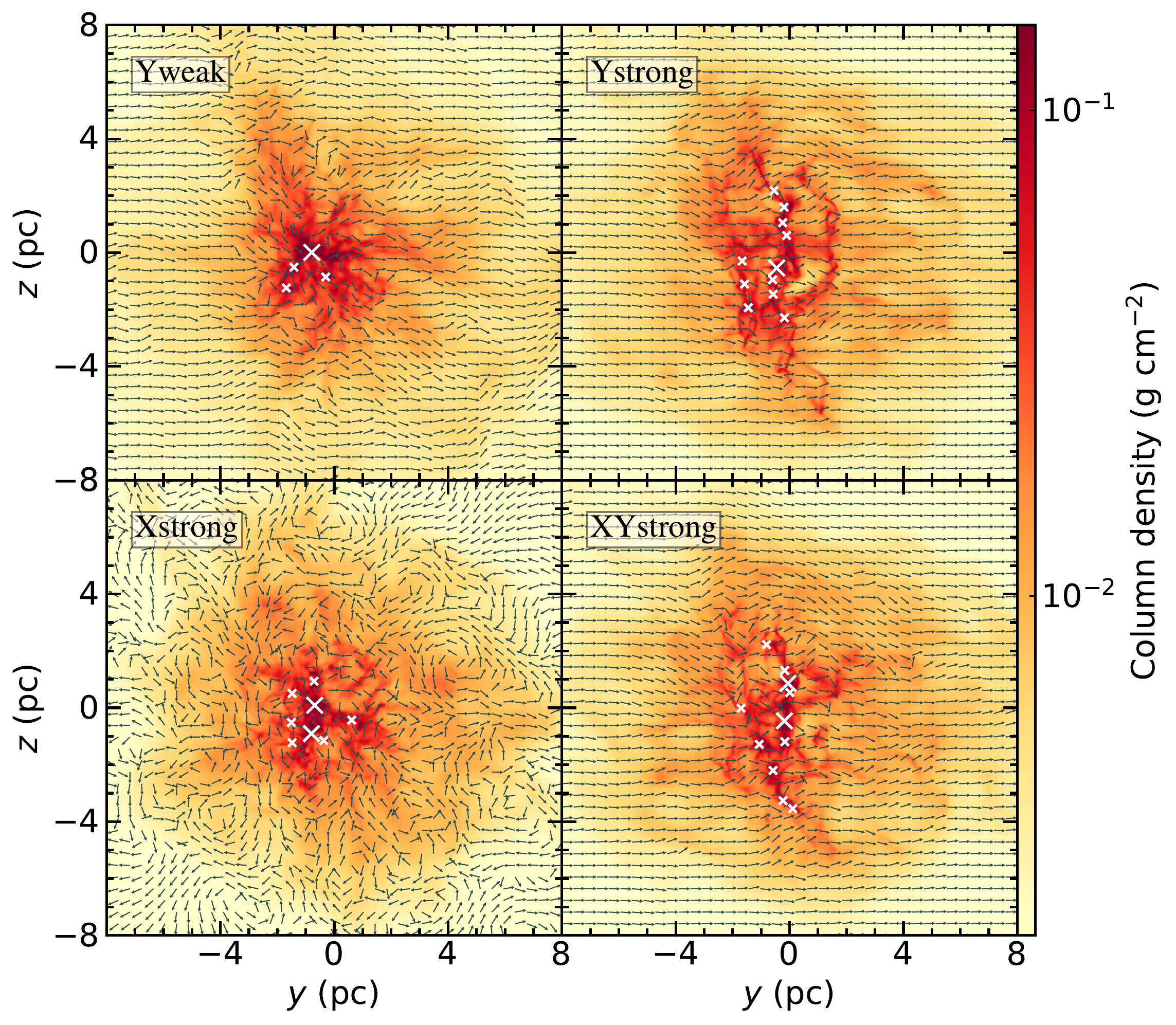}
    \end{center}
    \caption{The column density along the collision axis in the in the $y$-$z$ plane at $t$ = 3.0 Myr for the Yweak (top left), Ystrong (top right), Xstrong (bottom left), and XYstrong (bottom right) models. The large crosses show massive dense cores more than 100 $M_{\odot}$, and small crosses show dense cores of 10 $M_{\odot}$ $<$ $M_{\rm core}$ $<$ 100 $M_{\odot}$. The color bar shows the column density, and the arrows show normalized mass-weighted magnetic fields averaged along the collision axis.}
    \label{fig17} 
\end{figure*}

In figure \ref{fig17}, we show the position of cores with masses greater than 10 $M_{\odot}$ in the strong $B_0$ models, the Xstrong, Ystrong, and XYstrong models, and the weak $B_0$ model, the Yweak model, overlaid on the plot of column density looking from the collision axis ($x$-axis). Large crosses show cores with more than 100 $M_{\odot}$, and small crosses show cores with 10 $M_{\odot}$ $<$ $M_{\rm core}$ $<$ 100 $M_{\odot}$. These massive dense cores are distributed along the filaments with column densities greater than $10^{-1}$ g cm$^{-2}$ in the strong $B_0$ models. In the Ystrong and XYstrong models, the filaments hosting the massive dense cores are roughly perpendicular to the normalized mass-weighted magnetic field lines of which directions are shown by unit arrows in figure \ref{fig17}.

If the core mass function, $\phi$, is defined as
\begin{equation}
\phi = \frac{dN}{dM_{\rm core}} \propto M_{\rm core}^{-\gamma},
\end{equation}
where $M_{\rm core}$ is the core mass, $dN$ is the number of cores with masses between $M_{\rm core}$ and $M_{\rm core}$+$dM_{\rm core}$, and $\gamma$ is power index of core mass function, the cumulative core mass distribution, $N(\geq M_{\rm core})$, is given by,
\begin{equation}
N(\geq M_{\rm core})=\int_{M_{\rm core}}^{\infty} \frac{dN}{dM_{\rm core}}dM_{\rm core} \propto M_{\rm core}^{-(\gamma-1)}\propto M_{\rm core}^{\alpha} \label{cmf}
\end{equation}
where $\alpha$ = -($\gamma$-1). The least square fits ($\alpha$) using equation \ref{cmf} with standard deviation for cumulative mass distributions of cores with masses greater than 10 $M_{\odot}$ for all models at $t$ = 3.0 Myr are shown in figure \ref{fig:cmf-all} . The power indexes of core mass functions, $\gamma$, are $\gamma$ $\sim$ 1.3 - 1.4 in the weak $B_0$ models and $\gamma$ $\sim$ 1.5 - 1.9 in the strong $B_0$ models. The strong $B_0$ models have slightly larger $\gamma$ than that in the weak $B_0$ models.
These $\gamma$ values are similar to those of HD simulations performed by \citet{2014ApJ...792...63T} and \citet{2018PASJ...70S..58T}.
These values are also closer to the observed power indexes of core mass functions \citep{2007ApJ...665.1194I,2019ApJ...872..121U}, indicating the possible integral role of magnetic fields in determining the observed core mass function.
 
\section{Discussion}\label{discussion}

Our simulation results have shown that a greater number of massive dense cores ($>$ 10 $M_{\odot}$) form in the strong $B_0$ models than the weak $B_0$ models.
In the weak $B_0$ models, NTSI develops in the shocked layer produced by the CCC and induces the quasi-periodic shifts of the shocked layer in the early stage of CCC.   
Gas concentrations develop at the extremes of the shifts.
In these gas concentrations, dense cores of small mass form earlier than the strong $B_0$ models.
In the strong $B_0$ models, the turbulent magnetic fields suppress such small-scale NTSI. The turbulent magnetic fields increase the typical scale of NTSI and the thickness of the shocked layer. The turbulent magnetic fields also contribute to the increase in mass of dense cores in the thick shocked layer. Both effects contribute to the formation of massive dense cores in the strong $B_0$ models. 
The suppression effect of the magnetic field on NTSI in a shocked region is studied by {\citet{2007ApJ...665..445H}}. They have reported that magnetic fields parallel to the shock are more effective in suppression of the NTSI than magnetic fields normal to the shock. 
This effect can be the reason why the direction of $B_0$ to the collision axis affects the core mass functions.

\begin{figure*}
    \begin{center}
    \includegraphics[width=0.7\textwidth]{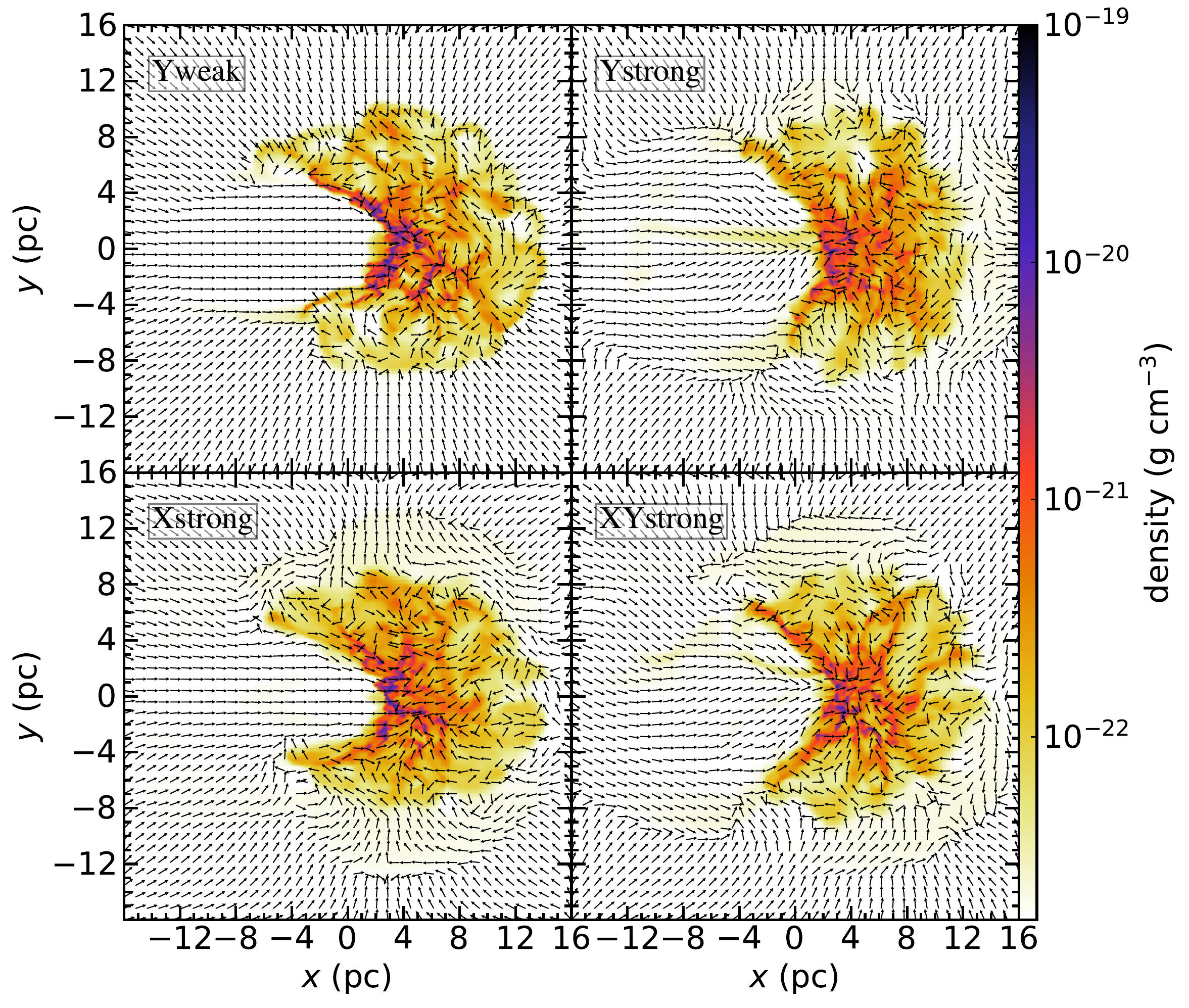}
    \end{center}
    \caption{Slice plots of the gas density in $z$ = 0 pc at $t$ = 2.0 Myr for the Yweak (top left), Ystrong (top right), Xstrong (bottom left), and XYstrong (bottom right) models. These panels show that the concave structures produced by the small cloud depend on the direction of $B_0$. The arrows show normalized vectors, [$v_x$/(${v_x}^2$ + ${v_y}^2$)$^{1/2}$, $v_y$/(${v_x}^2$ + ${v_y}^2$)$^{1/2}$]. Color bar on the right edge shows the gas density.}
    \label{fig15} 
\end{figure*}

In figure \ref{fig15}, we show the concave structures of the shocked layer created by the small cloud penetration into the large cloud in the strong $B_0$ models at $t$ = 2.0 Myr, as well as for the Yweak model for comparison (all shock fronts in the weak $B_0$ models look essentially identical). The concave structure indicates a converging flow to the collision axis for the post-shock gas of the small cloud in the left side part of the shocked layer and a diverging flow from the collision axis for the post-shock gas of the large cloud in the right-hand-side part of the shocked layer, since the appearance of oblique part of the concave structure implies that oblique shock wave is formed by the collision of the clouds. The converging flow of the post-shock gas of the small cloud accumulates gas in the shocked layer to the collision axis and contributes to the mass increase of the massive dense cores in the post-shock gas of the small cloud in the later stage of the collision. On the other hand, the diverging flow of the post-shock gas of the large cloud moves away from the collision axis and reduces the gas mass of the shocked layer. Stronger diverging flow appears at the shock more distant from the collision axis, as shown by normalized vectors, [$v_x$/(${v_x}^2$ + ${v_y}^2$)$^{1/2}$, $v_y$/(${v_x}^2$ + ${v_y}^2$)$^{1/2}$] in figure \ref{fig15}, although the diverging flow is highly disturbed by the turbulent flow in clouds. The diverging flow effect of reducing the gas mass of the shocked layer is weaker than the converging flow effect near the collision axis. After the left-hand-side shock of the shocked layer has swept up the small cloud at $t$ = 1.7 Myr, we expect a rarefaction wave to propagate rightward in the shocked layer. However, this rarefaction wave does not affect the dominance of the converging flow in the post-shock gas, since speeds of gas flows caused by the rarefaction wave are as high as the magnetosonic speed $\sim$ 0.4-3 km s$^{-1}$, and they are much less than the converging gas speeds $\sim$ 5-10 km s$^{-1}$ at $t$ = 2.0 Myr, which is 0.3 Myr after the complete penetration of the small cloud into the large cloud, as shown in figure \ref{fig15}. The converging gas flow is dominant in the shocked layer at $t$ = 2.0 Myr, as shown by the normalized velocity vectors in figure \ref{fig15}. The shape of the concave structure should be closely related to the strength of the converging flow. If the concave structure is widely opened, we expect a weak converging flow. If the concave structure is instead narrow, we expect a stronger converging flow. Figure \ref{fig15} shows that among the strong $B_0$ models the concave structure is most widely opened in the Ystrong model and is most narrow in the Xstrong model. The concave structure in the XYstrong is intermediate between them. We expect a stronger converging flow in the Xstrong and a weaker converging flow in the Ystrong model. We also expect that the converging flow in the XYstrong is intermediate between them. The concave structures in those models are consistent with the difference of core mass distributions, if the converging flow indeed contributes to the formation of massive dense cores. The concave structures in the weak $B_0$ models are similar to the Xstrong model. We expect rather strong converging flow in the weak $B_0$ models, and this can explain the reason why the mass of the most massive dense core is larger in the weak $B_0$ models than the strong $B_0$ models, as shown in figure \ref{fig:cmf-all}.

We discuss a possible role of magnetic field on self-gravitational instability in the shocked layer. The gravitational instability condition for a disk with a magnetic field $B_n$ is given by
\begin{equation}
\frac{\Sigma \sqrt{4\pi^2G}}{B_n}=16.2 \left(\frac{\Sigma}{0.01 \textrm{ g cm}^{-2}}\right)\left(\frac{B_n}{1\textrm{ } \mu\textrm{G}}\right)^{-1}>1,\label{gravinst1}
\end{equation}
where $B_n$ is the perpendicular magnetic field to the disk and $\Sigma$ is the surface density of the disk \citep{1978PASJ...30..671N,1983PASJ...35..187T}. A disk with a parallel magnetic field is gravitationally unstable for a perturbation with a wave number vector parallel to the magnetic field \citep{1983PASJ...35..187T,1998ApJ...506..306N}. We estimate the gravitational unstable scale of this case according to \citet{2014ApJ...785...24T}. In \citet{2014ApJ...785...24T}, the maximum mass per unit length, $\lambda_{\rm max}$, of a gravitational equilibrium filament perpendicular to the magnetic field $B$ is given as
\begin{equation}
\lambda_{\rm{max}} = 0.24\frac{B R_0}{\sqrt{G}}\label{gravinst2}
\end{equation}
for the limiting case of the magnetic field energy being much larger than the internal gas energy in the filament, where $R_0$ is the filament radius. The gravitational instability condition of a filament with a width $h$ and with surface density, $\Sigma$, is 
\begin{equation}
h > h_{\rm min} = \lambda_{\rm max}/\Sigma = 0.24\frac{B R_0}{\sqrt{G}\Sigma},\label{gravinst3}
\end{equation}
since $\lambda$ of this filament is given as $\lambda$ = $h\Sigma$. We estimate the averaged magnetic field, $\langle B \rangle$, and the averaged surface density, $\langle \Sigma \rangle$, of the shocked layer formed in our numerical simulations by using a thick disk region with a radius of 4 pc and with a thickness of 2.5 pc that can contain the shocked layer. In the Xstrong, Ystrong, and XYstrong models at $t$ = 1.7 Myr (a similar epoch at which the small cloud completely penetrates the large cloud), the averaged $\langle B_x \rangle$ = 11.8 $\mu$G, 16.4 $\mu$G, and 15.8 $\mu$G, the averaged $\langle B_y \rangle$ = 10.1 $\mu$G, 25.9 $\mu$G, and 21.4 $\mu$G, and the averaged surface density is $\langle \Sigma \rangle$ = 0.018 g cm$^{-2}$, 0.015 g cm$^{-2}$, and 0.016 g cm$^{-2}$, respectively. We note that $\langle B_x \rangle$ and $\langle B_y \rangle$ in the weak $B_0$ models are much smaller than in the strong $B_0$ models, and averaged surface density is $\langle \Sigma \rangle$ = 0.019-0.020 g cm$^{-2}$ in the weak $B_0$ models. Using these values, from equation (\ref{gravinst1}), we find that $\langle B_x \rangle$ cannot suppress gravitational instability of the shocked layers in the strong $B_0$ models.
Using $\langle B_y \rangle$ and $\langle \Sigma \rangle$, from equation (\ref{gravinst3}) we get $h_{\rm min}$ as 0.65 pc, 2.0 pc, and 1.6 pc for the Xstrong, Ystrong, and XYstrong models, respectively.
These values of $h_{\rm min}$ are less than the typical, lateral size ($\sim$ 8 pc) of the shocked layers. We also estimate gravitational instability of the filaments formed in our models. We find five filaments in shocked layers, one filament in each model expect for the Ystrong model, at $t$ = 1.7 Myr, and we estimate their $\lambda$/$\lambda_{\rm max}$. The width of filaments in the Xstrong, XYstrong, Xweak, Yweak, and XYweak are 0.2, 0.6 pc, 0.3 pc, 0.3 pc, and 0.3 pc, respectively, and their line masses, $\lambda$, estimated using surface density and the width of the filaments are 16 $M_{\odot}$ pc$^{-1}$, 48 $M_{\odot}$ pc$^{-1}$, 30 $M_{\odot}$ pc$^{-1}$, 28 $M_{\odot}$ pc$^{-1}$, and 28 $M_{\odot}$ pc$^{-1}$, respectively. The critical line masses, $\lambda_{\rm max}$, for these filaments are estimated from the magnetic field threading the filaments and the width of the filaments, and the $\lambda$/$\lambda_{\rm max}$ for these filaments are 4.6, 1.6, 26.7, 8.9, and 5.9, respectively. Since $\lambda_{\rm max}$ $<$ $\lambda$, the filaments are gravitationally unstable. This suggests that the shocked layers are gravitationally unstable in all models and that a typical mass of a fragment formed by the gravitational instability is $\Sigma h_{\rm min}^2$, which is much larger than masses of dense cores formed in the early stage of the collisions of clouds. From the discussion, the main effects of magnetic fields on the CCCs are the suppression of the NTSI and the increase of the thickness of the shocked layers formed in the strong $B_0$ models, as shown in section \ref{results}.

We estimate the magnetic field strength $B_0$ that can suppress NTSI in the shocked layer from our simulation results. 
We can expect a magnetic field suppression effect on the NTSI, if 
\begin{equation}
   \frac{B^2}{8\pi \lambda} > \rho \frac{(\Delta v)^2}{\lambda}, \label{Bsuppress}
\end{equation}
where $B$ and $\rho$ are magnetic field strength and gas density, respectively, in the shocked layer, $\lambda$ is the typical scale of the NTSI, and $\Delta v$ is the perturbed velocity induced by the NTSI. 
If the magnetic pressure enhanced by the shock compression is dominant in the shocked layer, we can expect
\begin{equation}
   \rho_0 v_{\rm sh}^2 \sim {B^2}/{8\pi}, \label{shockcond1}
\end{equation}
\begin{equation}
   \rho = \rho_0 B/(\alpha B_{0}), \label{shockcond2}
\end{equation}
and 
\begin{equation}
   \rho v = \rho_0 v_{\rm sh} \label{shockcond3}
\end{equation}
\begin{figure*}
    \begin{center}
    \includegraphics[width=0.7\textwidth]{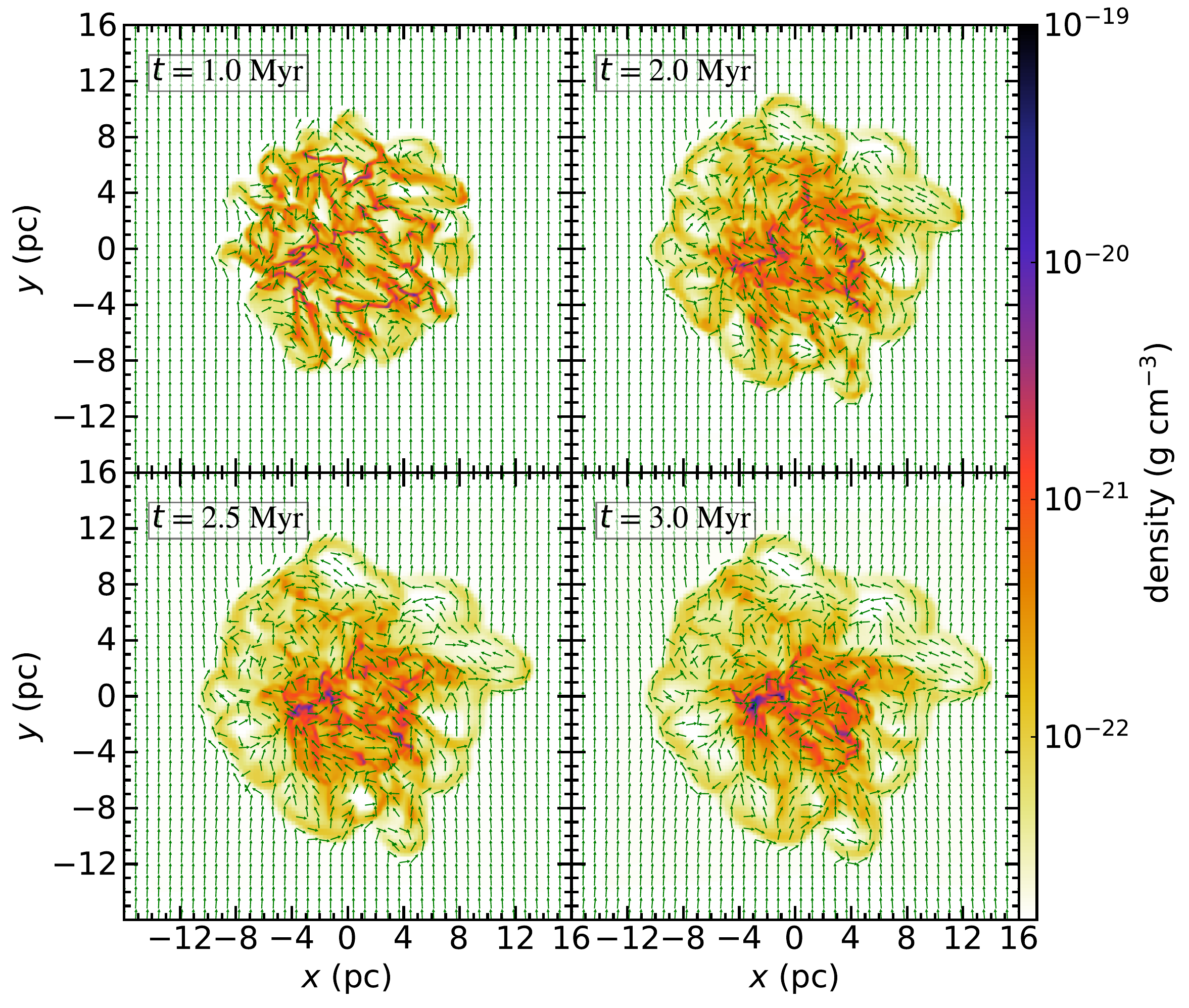}
    \end{center}
    \caption{Same as figure \ref{fig:yweak5}, but for the ISweak model.}
    \label{fig:isweak} 
\end{figure*}
\begin{figure*}
    \begin{center}
    \includegraphics[width=0.7\textwidth]{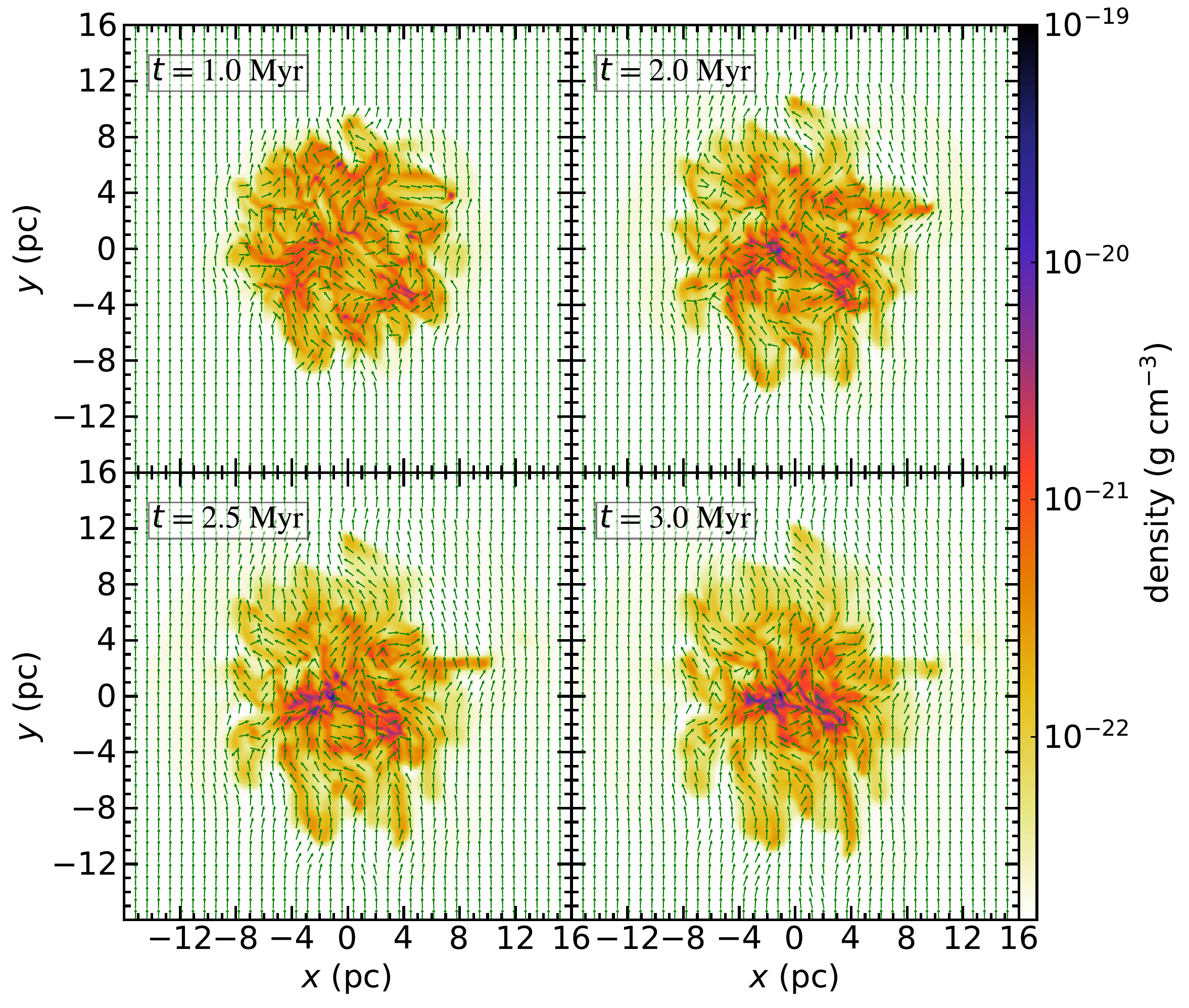}
    \end{center}
    \caption{Same as figure \ref{fig:yweak5}, but for the ISstrong model.}
    \label{fig:isstrong} 
\end{figure*}
from the MHD shock-wave condition in the rest frame of the shock front, where $v_{\rm sh}$ and $v$ are pre-shock and post-shock gas velocities, respectively, and $\alpha B_{0}$ and $\rho_0$ are pre-shock parallel component of the magnetic field to the shock front and gas density, respectively. Here, $\alpha$ is an enhancement factor of the parallel component of the magnetic field to the shock front induced by the turbulent motion in the clouds before the collision.

The suppression condition of NTSI can be given as
\begin{equation}
B_{0}>\frac{\sqrt{8\pi\rho_0}v_{\rm sh}}{\alpha}\left(\frac{\Delta v}{v_{\rm sh}}\right)^2    
\end{equation}
from equation (\ref{Bsuppress}), using equation (\ref{shockcond1}) and equation (\ref{shockcond2}).
If ${\Delta v}/{v_{\rm sh}}\sim 0.2$ and $\alpha \sim 1$, we have
\begin{equation}
  B_{0}> 1.92 \left(\frac{\rho_0}{3.67\times 10^{-22} \textrm{ g cm}^{-3}}\right)^{0.5} \left(\frac{v_{\rm sh}}{5 \textrm{ km s}^{-1}}\right)\mu\textrm{G}.\label{Bsuppress1}  
\end{equation}
Here, we assume that $v_{\rm sh}$ is roughly half of the collision speed.  
This estimated value is consistent with our numerical results, since the strong $B_0$ is much larger than this value and the weak $B_0$ is much less than this value.

If the magnetic field pressure is dominant in the shocked layer compared to the effects of the turbulent motions and the thermal gas, we estimate magnetic field strength, $B$, and gas density, $\rho$, in the shocked layer as,
\begin{equation}
B=48.0 \left(\frac{\rho_0}{3.67\times 10^{-22} \textrm{ g cm}^{-3}}\right)^{0.5}\left(\frac{v_{\rm sh}}{5 \textrm{ km s}^{-1}}\right)\mu\textrm{G} 
\end{equation}
and
$$
\hspace{-1.99cm}\rho= 4.39\times 10^{-21} \left(\frac{\rho_0}{3.67\times 10^{-22} \textrm{ g cm}^{-3}}\right)^{1.5}
$$
\begin{equation}
\hspace{0.6cm}\times\left(\frac{v_{\rm sh}}{5 \textrm{ km s}^{-1}}\right)\left(\frac{\alpha B_{0}}{4\textrm{ }\mu\textrm{G}}\right)^{-1} \textrm{g cm}^{-3}
\end{equation}
using equation (\ref{shockcond1}) and equation (\ref{shockcond2}).

After the whole small cloud penetrates the shocked layer, the shocked layer will change its structure in the free-fall timescale of the shocked layer $\sim$ 1 Myr, since the ram pressure by the small cloud gas does not push the shocked layer after the whole small cloud penetrates the large cloud. In this stage, dense core formation and core mass evolution proceed to form massive bound cores in the timescale of $t_{\rm ff}$, as shown in subsection \ref{strong}. We thus propose that if the shocked layer moves out of the large cloud in less than $t_{\rm ff}$ after the whole small cloud penetrates, then such massive dense core formation will not proceed. 
We have adopted a collision speed of 10 km s$^{-1}$ in this study.
We briefly discuss the consequences of a higher collision speed, since the observed collision speeds are in the range of 10 to 20 km s$^{-1}$ \citep{2018PASJ...70S..46F}.
If we assume the collision speed of 20 km s$^{-1}$, $v_{\rm sh}$ = 10 km s$^{-1}$, then the condition of suppression of NTSI will be 
\begin{equation}
B_0 > 3.84 \left(\frac{\rho_0}{3.67\times 10^{-22} \textrm{ g cm}^{-3}}\right)^{0.5} \left(\frac{v_{\rm sh}}{10 \textrm{ km s}^{-1}}\right)\mu\textrm{G}
\end{equation}
from equation (\ref{Bsuppress1}).
This result indicates that stronger $B_0$ than that used in this study is required to suppress the NTSI.  
The free-fall time of the shocked layer is $t_{\rm ff} \sim 0.7 \textrm{ Myr}$ for the collision speed of 20 km s$^{-1}$.
If the penetration time of the small cloud is 
\begin{equation}
t_{\rm penetration}= \frac{2R_{\rm small}}{v_{\rm sh}}\sim 0.7 \textrm{ Myr}
\end{equation}
and the crossing time of the small cloud to the large cloud is
\begin{equation}
t_{\rm cross}=\frac{2R_{\rm large}}{v_{\rm sh}}\sim 1.4 \textrm{ Myr},
\end{equation}
$t_{\rm cross}$ is comparable to sum of $t_{\rm penetration}$ and $t_{\rm ff}$.
This means that larger cloud sizes of the large cloud are needed to induce massive dense core formation. 
In our galaxy, various sizes and masses of GMCs are observed.
Magnetic field strength depends on location in our galaxy \citep{2015A&ARv..24....4B}.
We will extend our study to a higher collision speed case with larger cloud sizes and stronger magnetic fields in our future works. We will study protostar formation and stellar feedback effects on massive star formation by CCCs, using sink particles in our future works.

\section{Summary}\label{summary}
We have performed magnetohydrodynamic simulations of the cloud-cloud collision to study the role of the magnetic field on massive dense core formation in the colliding clouds. We selected two clouds with masses of 972 $M_{\odot}$ and 7774 $M_{\odot}$ with the typical density of giant molecular clouds and with internal turbulence such that the clouds are in the virial equilibrium. Two cases of uniform magnetic field strengths, $B_0$ = 4.0 $\mu$G (strong) and 0.1 $\mu$G (weak), and three cases of uniform magnetic field directions, parallel, perpendicular, and oblique to collision axis, were studied. Magnetic fields were modified by internal turbulent motion in the clouds. The distribution of magnetic field strength and gas density in the clouds in the strong $B_0$ model is consistent with the relation observed by \citet{2010ApJ...725..466C}. The small cloud is given a collision speed of 10 km s$^{-1}$ after the turbulent magnetic field generation in the clouds.
We have also simulated the evolution of the isolated clouds with the same uniform magnetic field strengths as in colliding clouds for comparison. Our main conclusions are as follows.

1. In the weak $B_0$ models, quasi-periodic shifts with small size appear in the shocked layer formed by cloud-cloud collision and develop with time. The quasi-periodic spatial shifts should be caused by nonlinear thin shell instability. Dense cores are formed at the extremes of the shifts. 
In the strong $B_0$ models, such shifts are suppressed by the stronger magnetic field, and a greater number of massive dense cores are formed than in the weak $B_0$ models. The number of massive bound cores in which self-gravitational energy dominates over turbulent energy and magnetic field energy is also larger than the weak $B_0$ models.
In the massive bound cores with more than 10 $M_{\odot}$, we can expect massive star formation, since the free-fall time of these cores is less than 0.3 Myr.
In isolated cloud models, the bound cores form earlier and are less massive than the colliding clouds models. 
Since their masses are less than 3 $M_{\odot}$ and their free-fall times are less than 0.3 Myr, we can expect only intermediate-mass star formation in these cores.
 
2. The cumulative mass distributions of dense cores formed in our simulation models clearly show that a greater number of massive dense cores are formed in the strong $B_0$ models than in the weak $B_0$ models. 

3. In the strong $B_0$ models, massive dense cores distribute in dense gas filaments of which directions are roughly normal to the direction of $B_0$ except for the Xstrong (strong $B_0$ parallel to collision axis) model.

4. We give a simple analytic model for the magnetic field strength needed to suppress the instability of the shocked layer formed by colliding clouds and thus suppress low mass core formation. The magnetic field strength is related to collision speed and cloud size. Testing this model further will be the subject of future works.

\begin{ack}
The authors thank Yasuo Fukui, Elizabeth Tasker, Kazuyoshi Torii, Tsuyoshi Inoue, Kazuo Sorai, and Syu-ichiro Inutsuka for their fruitful discussions. Numerical computations were carried out on the Cray XC30 and XC50 supercomputer systems at the Center for Computational Astrophysics of the National Astronomical Observatory of Japan. Numerical analysis was done using the yt, a data analysis and visualization package \citep{2011ApJS..192....9T}. NS is financially supported by the MEXT scholarship. AH is funded by the JSPS KAKENHI Grant (JP19K03923). TO is financially supported by MEXT KAKENHI Grant (18H04333) and the Grant-in-Aid (19H01931).
\end{ack}

\appendix 
\section*{Isolated cloud models}\label{app}
\begin{figure}
    \begin{center}
    \includegraphics[width=0.5\textwidth]{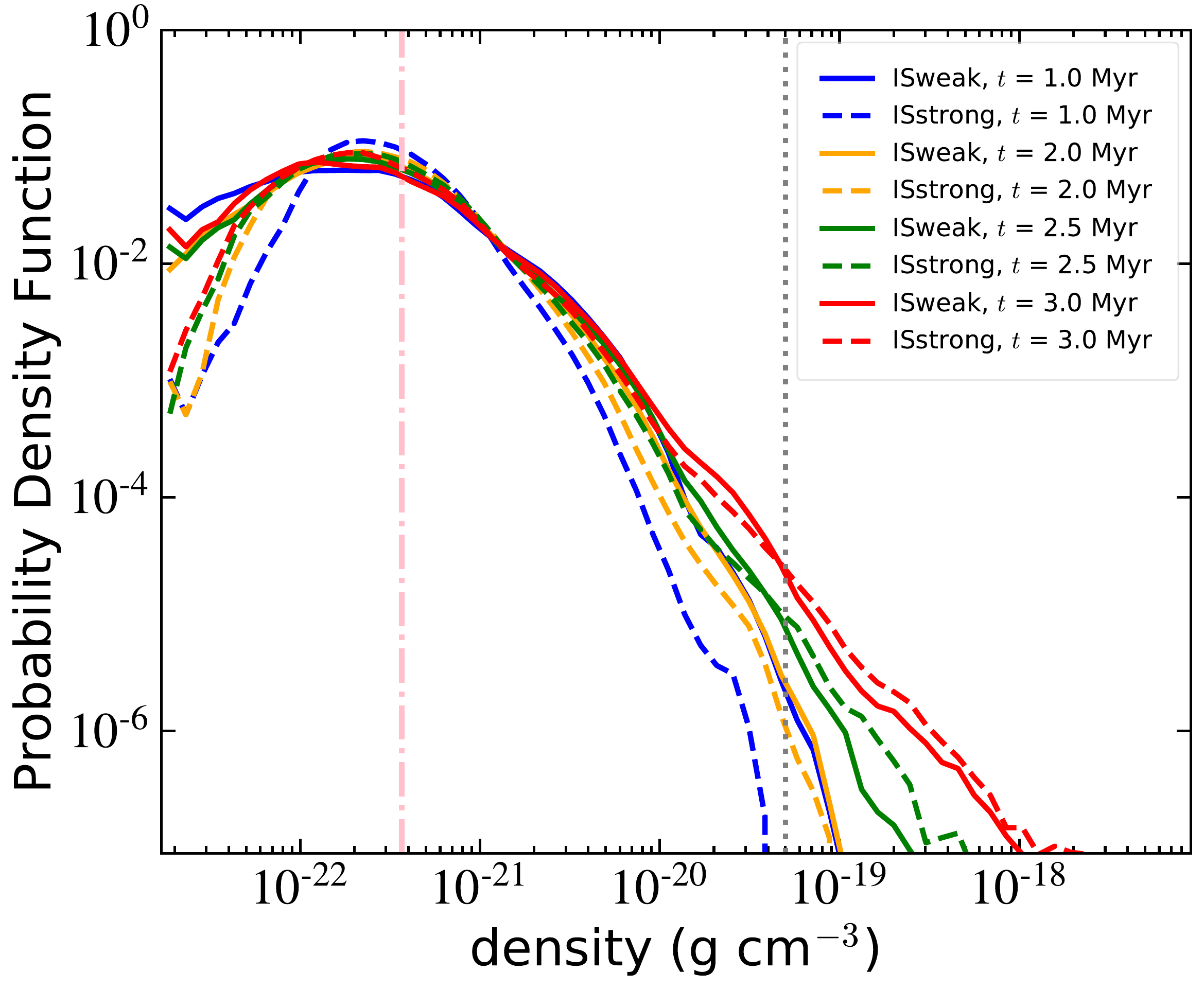}
    \end{center}
    \caption{Probability density functions (PDFs) of the isolated clouds at $t$ = 1.0 Myr (blue lines), 2.0 Myr (orange lines), 2.5 Myr (green lines), and 3.0 Myr (red lines) for the ISweak (solid lines) and ISstrong (dashed lines) models. The vertical, dash-dotted, and dotted lines indicate the initial density of the cloud and the density threshold, $\rho_{\rm th}$, for dense cores, respectively. The selection criteria for volumes used for PDFs is same as in figure \ref{fig:3}.}
    \label{fig:ispdf} 
\end{figure}
We show results of the isolated cloud models (ISweak and ISstrong models). The density slice plots at $t$ = 1.0 Myr, 2.0 Myr, 2.5 Myr, and 3.0 Myr for the ISweak model ($B_0$ = 0.1 $\mu$G) are shown in figure \ref{fig:isweak}. The initial uniform magnetic field is distorted by the turbulence. The gas density contrast is clearly seen at $t$ = 1.0 Myr. Dense regions are formed at later epochs ($t$ = 2.5 Myr and 3.0 Myr) due to gas motion induced by self-gravity of gas. The density slice plots at $t$ = 1.0 Myr, 2.0 Myr, 2.5 Myr, and 3.0 Myr for the ISstrong model ($B_0$ = 4.0 $\mu$G) are shown in figure \ref{fig:isstrong}. Contrary to the ISweak model, the gas in the ISstrong model has less density contrast at $t$ = 1.0 Myr, since the strong magnetic field suppresses density enhancement by the turbulence. At $t$ = 2.0 and 2.5 Myr, more gas accumulation towards $x$-$z$ plane is seen in the ISstrong model than the ISweak model. This is because the strong magnetic field induces more gas flow along the magnetic field lines than the weak magnetic field. At $t$ = 3.0 Myr, we find the gas flow by the self-gravity towards dense gas regions in both models. We find more gas concentration near the $x$-$z$ plane region at $t$ = 3.0 Myr in the ISstrong model than the ISweak model. PDFs of the isolated clouds at $t$ = 1.0 Myr, 2.0 Myr, 2.5 Myr, and 3.0 Myr for the ISweak and ISstrong models are shown in figure \ref{fig:ispdf}. In figure \ref{fig:ispdf}, the initial density of the isolated cloud and density threshold, $\rho_{\rm th}$, for dense cores are indicated by the
vertical, dash-dotted, and dotted lines, respectively. The PDFs at $t$ = 1.0 Myr show higher gas density contrast in the ISweak model than the ISstrong model. The PDFs at later epochs ($t$ = 2.5 Myr and 3.0 Myr) show power-law tail due to the effect of self-gravity and the formation of dense gas with a density greater than $\rho_{\rm th}$ in the ISweak and ISstrong models.

\bibliography{CCC.bib}

\end{document}